\documentclass[12pt]{article}
\usepackage[margin=1.1in]{geometry}
\usepackage{graphicx}
\usepackage{float}

\usepackage{enumerate}
\usepackage{amsmath}
\usepackage{amssymb}
\usepackage{amsthm}
\usepackage{color}
\usepackage{xcolor}
\usepackage{graphicx}
\usepackage{bbm}
\usepackage{color}
\usepackage{geometry}
\usepackage{bbding}
\usepackage{ dsfont }
\usepackage{caption}
\usepackage{hyperref}
\usepackage{stackrel}
\usepackage{graphicx,xcolor}
\usepackage{tikz}
\usepackage{hyperref}
\usepackage{enumerate}
\usepackage{multicol}
\usepackage{soul}
\usepackage{tabularx}
\usepackage{multirow}
\usepackage{mdwlist}
\usepackage{enumitem}
\usepackage{accents}
\usepackage{mathtools}
\usepackage{subcaption}
\makeatletter
\DeclareRobustCommand\widecheck[1]{{\mathpalette\@widecheck{#1}}}
\def\@widecheck#1#2{%
    \setbox\z@\hbox{\m@th$#1#2$}%
    \setbox\tw@\hbox{\m@th$#1%
       \widehat{%
          \vrule\@width\z@\@height\ht\z@
          \vrule\@height\z@\@width\wd\z@}$}%
    \dp\tw@-\ht\z@
    \@tempdima\ht\z@ \advance\@tempdima2\ht\tw@ \divide\@tempdima\thr@@
    \setbox\tw@\hbox{%
       \raise\@tempdima\hbox{\scalebox{1}[-1]{\lower\@tempdima\box
\tw@}}}%
    {\ooalign{\box\tw@ \cr \box\z@}}}
\makeatother

\usepackage{threeparttable}

\theoremstyle{empty}

\newtheorem{proposition}{Proposition}
\newtheorem{theorem}{Theorem}
\newtheorem{lemma}{Lemma}

\newcommand{\E}{\mathcal{E}}
\newcommand{\D}{D}
\newcommand{\e}{e}

\renewcommand{\(}{\left(}
\renewcommand{\)}{\right)}
\newcommand{\deleq}{\stackrel{\Delta}{=}}

\newcommand{\Sa}{S_a}
\newcommand{\Sd}{S_d}
\newcommand{\sa}{s_a}
\newcommand{\sd}{s_d}

\newcommand{\sigs}{\widetilde{\sigma}}
\newcommand{\sigas}{\widetilde{\sigma}_a}
\newcommand{\sigds}{\widetilde{\sigma}_d}
\newcommand{\sign}{\sigma}
\newcommand{\sigan}{\sign_a}
\newcommand{\sigdn}{\sign_d}
\newcommand{\sigaswe}{\widetilde{\sigma}_a^*}
\newcommand{\sigdswe}{\widetilde{\sigma}_d^*}
\newcommand{\siganwe}{\sigma_a^*}
\newcommand{\sigdnwe}{\sigma_d^*}
\newcommand{\pee}{\rho_e}
\newcommand{\ceq}{C}
\newcommand{\ua}{u_a}
\newcommand{\jhat}{\hat{j}}
\newcommand{\ud}{u_d}
\newcommand{\Ua}{U_a}
\newcommand{\Udzero}{U^0_d}
\newcommand{\Uazero}{U^0_a}

\newcommand{\signwe}{\sigma^*}
\newcommand{\sigswe}{\widetilde{\sigma}^*}

\newcommand{\Ud}{U_d}

\newcommand{\regimel}{L}
\newcommand{\regimem}{M}
\newcommand{\regimeh}{H}

\newcommand{\cd}{p_d}

\newcommand{\Ebar}{\bar{\E}}
\newcommand{\Czero}{C_{\emptyset}}
\newcommand{\Cs}{C_\s}
\newcommand{\Ce}{C_{\e}}
\newcommand{\ca}{p_a}

\newcommand{\psewe}{\pswe_\e}
\newcommand{\pebarss}{\psss_{\ebar}}
\newcommand{\pebarssp}{\psss_{\ebar}'}
\newcommand{\gamen}{\Gamma}
\newcommand{\games}{\widetilde{\Gamma}}

\newcommand{\gamezero}{\Gamma^0}
\newcommand{\cdij}{\cd^{ij}}
\newcommand{\cdtil}{\widetilde{p}_d}
\newcommand{\cdtilinv}{\widetilde{p}_d^{-1}}
\newcommand{\typeone}{\widetilde{\mathrm{I}}}
\newcommand{\typetwo}{\widetilde{\mathrm{II}}}

\newcommand{\ve}{v_e}

\newcommand{\Ebarp}{K}

\newcommand{\nnk}{E_{(k)}}
\newcommand{\Cpi}{C_{(k)}}
\newcommand{\regimei}{\Lambda^i}
\newcommand{\pess}{\tilde{\rho}_\e}
\newcommand{\pswe}{\tilde{\rho}^{*}}
\newcommand{\Erho}{\Ebar^{*}}
\newcommand{\regimeij}{\Lambda^i_j}
\newcommand{\rhoone}{\widetilde{P}}

\newcommand{\psssone}{\tilde{\rho}^\dagger}

\newcommand{\regimej}{\Lambda_j}

\newcommand{\Udi}{\Ud^{\regimei}}
\newcommand{\Udj}{\Ud^{\regimej}}
\newcommand{\Uai}{\Ua^{\regimei}}
\newcommand{\cdbar}{\bar{p}_d(\ca)}

\newcommand{\Uaj}{\Ua^{\regimej}}

\newcommand{\ep}{e^{'}}
\newcommand{\Cpk}{C_{(k)}}

\newcommand{\Ebari}{\Ebar_{(k)}}

\newcommand{\siganp}{\sigan^{'}}

\newcommand{\porderm}{\rho_{(m)}}
\renewcommand{\i}{i}

\newcommand{\Uas}{\widetilde{U}_a}

\newcommand{\Ehat}{\widehat{\E}}
\newcommand{\Edag}{\widecheck{\E}}

\newcommand{\Uds}{\widetilde{U}_d}

\newcommand{\sdp}{\sd^{'}}
\newcommand{\Sdbar}{\bar{S}_d}

\newcommand{\pd}{\rho(\sigdn)}

\newcommand{\ped}{\rho_\e(\sigdn)}

\newcommand{\pwe}{\rho^{*}}
\newcommand{\pewe}{\rho_\e^{*}}

\newcommand{\ebar}{\bar{\e}}

\renewcommand{\t}{t}

\newcommand{\T}{T}
\newcommand{\pbar}{\widehat{\rho}}
\newcommand{\thetat}{\theta^t}

\newcommand{\sran}{\mathbf{\s}}

\newcommand{\pebar}{\widehat{\rho}_\e}

\newcommand{\pessp}{\pess'}
\newcommand{\psssp}{\psss'}
\newcommand{\pebarbar}{\widehat{\rho}_{\ebar}}

\newcommand{\cae}{p_{a,\e}}
\newcommand{\cde}{p_{d, \e}}

\newcommand{\epe}{\epsilon_\e}

\newcommand{\ebarp}{\widehat{\e}}

\newcommand{\Emax}{\Ebar^{\diamond}}
\newcommand{\ebarnew}{\bar{e}}
\newcommand{\porderi}{\rho_{(i)}}
\newcommand{\porderipone}{\rho_{(i+1)}}
\newcommand{\porderone}{\rho_{(1)}}

\newcommand{\Siganwe}{\Sigma^{*}_a}
\newcommand{\Sigdnwe}{\Sigma^{*}_d}

\newcommand{\ps}{\theta}

\newcommand{\regimesj}{\widetilde{\Lambda}_j}
\newcommand{\regimesi}{\widetilde{\Lambda}^i}
\newcommand{\psss}{\tilde{\rho}}

\newcommand{\s}{s}
\renewcommand{\S}{S}

\newcommand{\thetazero}{\theta^0}

\begin{document}
\newtheorem{assumption}{Assumption}

\title{\LARGE \bf Securing Infrastructure Facilities: When does proactive defense help?
}


\author{Manxi Wu, and Saurabh Amin
\thanks{M. Wu is with the Institute for Data, Systems, and Society, and S. Amin is with the Department of Civil and Environmental Engineering, Massachusetts Institute of Technology (MIT), Cambridge, MA, USA
        {\tt\small \{manxiwu,amins\}@mit.edu}}%
}
\date{}

\maketitle

\begin{abstract}
Infrastructure systems are increasingly facing new security threats due to the vulnerabilities of cyber-physical components that support their operation. In this article, we investigate how the infrastructure operator (defender) should prioritize the investment in securing a set of facilities in order to reduce the impact of a strategic adversary (attacker) who can target a facility to increase the overall usage cost of the system. We adopt a game-theoretic approach to model the defender-attacker interaction and study two models: normal-form game -- where both players move simultaneously; and sequential game -- where attacker moves after observing the defender's strategy. For each model, we provide a complete characterization of how the set of facilities that are secured by the defender in equilibrium vary with the costs of attack and defense. Importantly, our analysis provides a sharp condition relating the cost parameters for which the defender has the first mover advantage. Specifically, we show that to fully deter the attacker from targeting any facility, the defender needs to proactively secure all ``vulnerable facilities'' at an appropriate level of effort. We illustrate the outcome of the attacker-defender interaction on a simple transportation network. We also suggest a dynamic learning setup to understand how this outcome can affect the ability of imperfectly informed users to make their decisions about using the system in the post-attack stage.  

\end{abstract}

\section{Introduction}
In this article, we consider the problem of strategic allocation of defense effort to secure one or more facilities of an infrastructure system that is prone to a targeted attack by a malicious adversary. The setup is motivated by the recent incidents and projected threats to critical infrastructures such as transportation, electricity, and urban water networks (\cite{moteff2004critical}, \cite{rinaldi2001identifying}, \cite{sandberg2015cyberphysical}, and \cite{leetaru_2015}). Two of the well-recognized security concerns faced by infrastructure operators are: (i) How to prioritize investments among facilities that are heterogeneous in terms of the impact that their compromise can have on the overall efficiency (or usage cost) of the system; and (ii) Whether or not an attacker can be fully deterred from launching an attack by proactively securing some of the facilities. Our work addresses these questions by focusing on the most basic form of strategic interaction between the system operator (defender) and an attacker, modeled as a normal form (simultaneous) or a sequential (Stackelberg) game. The normal form game is relevant to situations in which the attacker cannot directly observe the chosen security plan, whereas the sequential game applies to situations where the defender proactively secures some facilities, and the attacker can observe the defense strategy. 

In recent years, many game-theoretical models have been proposed to study problems in cyber-physical security of critical infrastructure systems; see \cite{alpcan2010network}, and \cite{manshaei2013game} for a survey of these models. These models are motivated by the questions of strategic network design (\cite{dziubinski2013network}, \cite{laporte2010game} and \cite{schwartz2011network}), intrusion detection (\cite{chen2009game}, \cite{alpcan2003game}, \cite{dritsoula2012game}, and \cite{sethi2017value}), interdependent security (\cite{nguyen2009stochastic}, and \cite{amin2013security}), network interdiction (\cite{washburn1995two}, and \cite{dahan2015network}), and attack-resilient estimation and control (\cite{cardenas2011attacks}, and \cite{sridhar2012cyber}).

Our model is relevant for assessing strategic defense decisions for an infrastructure system viewed as a collection of facilities. In our model, each facility is considered as a distinct entity for the purpose of investment in defense, and multiple facilities can be covered by a single investment strategy. The attacker can target a single facility and compromise its operation, thereby affecting the overall operating efficiency of the system. Both players choose randomized strategies. The performance of the system is evaluated by a usage cost, whose value depends on the actions of both players. In particular, if an undefended facility is targeted by the attacker, it is assumed to be compromised, and this outcome is reflected as a change in the usage cost. Naturally, the defender aims to maintain a low usage cost, while the attacker wishes to increase the usage cost. The attacker (resp. defender) incurs a positive cost in targeting (resp. securing) a unit facility. Thus, both players face a trade-off between the usage cost and the attack/defense costs, which results in qualitatively different equilibrium regimes.

We analyze both normal form and sequential games in the above-mentioned setting. First, we provide a complete
characterization of the equilibrium structure in terms of the relative vulnerability of different facilities and the costs of defense/attack for both games. Secondly, we identify ranges of attack and defense costs for which the defender gets the first mover advantage by investing in proactive defense. Furthermore, we relate the outcome of this game (post-attack stage) to a dynamic learning problem in which the users of the infrastructure system are not fully informed about the realized security state (i.e. the identity of compromised facility).   

We now outline our main results. To begin our analysis, we make the following observations. Analogous to \cite{dritsoula2017game}, we can represent the defender's mixed strategy by a vector with elements corresponding to the probabilities for each facility being secured. The defender's mixed strategy can also be viewed as her effort on each facility. Moreover, the attacker/defender only targets/secures facilities whose disruption will result in an increase in the usage cost (Proposition \ref{strict_dominated}). If the increase in the usage cost of a facility is larger than the cost of attack, then we say that it is a vulnerable facility.

Our approach to characterizing Nash equilibrium (NE) of the normal form game is based on the fact that it is strategically equivalent to a zero-sum game. Hence, the set of attacker's equilibrium strategies can be obtained as the optimal solution set of a linear optimization program (Proposition \ref{opt_eq}). For any given attack cost, we show that there exists a threshold cost of defense, which distinguishes two equilibrium regime types, named as type I and type II regimes. Theorem \ref{attacker_strategy} shows that when the defense cost is lower than the cost threshold (type I regimes), the total attack probability is positive but less than 1, and all vulnerable facilities are secured by the defender with positive probability. On the other hand, when the defense cost is higher than the threshold (type II regimes), the total attack probability is 1, and some vulnerable facilities are not secured at all. 

We develop a new approach to characterize the subgame perfect equilibrium (SPE) of the sequential game, noting that the strategic equivalence to zero-sum game no longer holds in this case. In this game, the defender, as the first mover, either proactively secures all vulnerable facilities with a threshold security effort so that the attacker does not target any facility, or leaves at least one vulnerable facility secured with an effort less than the threshold while the total attack probability is 1. For any attack cost, we establish another threshold cost of the defense, which is strictly higher than the corresponding threshold in the normal form game. This new threshold again distinguishes the equilibrium strategies into two regime types, named as type $\typeone$ and type $\typetwo$ regimes. Theorem \ref{theorem:SPE} shows that when the defense cost is lower than the cost threshold (type $\typeone$ regimes), the defender can fully deter the attacker by proactively securing all vulnerable facilities with the threshold security effort. On the other hand, when the defense cost is higher than the threshold (type $\typetwo$ regimes), the defender maintains the same level of security effort as that in NE, while the total attack probability is 1. 

Our characterization shows that both NE and SPE satisfy the following intuitive properties: (i) Both the defender and attack prioritize the facilities that results in a high usage cost when compromised; (ii) The attack and defense costs jointly determine the set of facilities that are targeted or secured in equilibrium. On one hand, as the attack cost decreases, more facilities are vulnerable to attack. On the other hand, as the defense cost decreases, the defender secures more facilities with positive effort, and eventually when the defense cost is below a certain threshold (defined differently in each game), all vulnerable facilities are secured with a positive effort; (iii) Each player's equilibrium payoff is non-decreasing in the opponent's cost, and non-increasing in her own cost.

It is well-known in the literature on two player games that so long as both players can choose mixed strategies, the equilibrium utility of the first mover in a sequential game is no less than that in a normal form game (\cite{bacsar1998dynamic} (pp. 126), \cite{von2004leadership}). However, cases can be found where the first mover advantage changes from positive to zero when the attacker's observed signal of the defender's strategy is associated with a noise (\cite{bagwell1995commitment}). 
In the security game setting, the paper \cite{bier2007choosing} analyzed a game where there are two facilities, and the attacker's valuation of each facility is private information. They identify a condition under which the defender's equilibrium utility is strictly higher when his strategy can be observed by the attacker. In contrast, our model considers multiple facilities, and assumes that both players have complete information of the usage cost of each facility. 

In fact, for our model, we are able to provide sharp conditions under which proactive defense strictly increases the defender's utility. Given any attack cost, unless the defense cost is ``relatively high'' (higher than the threshold cost in the sequential game), proactive defense is advantageous in terms of strictly improving the defender's utility and fully deterring the attack. However, if the defense cost is ``relatively medium'' (lower than the threshold cost in sequential game, but higher than that in the normal form game), a higher security effort on each vulnerable facility is required to gain the first mover advantage. Finally, if the defense cost is ``relatively low'' (lower than the threshold cost in the normal form game), then the defender can gain advantage by simply making the first move with the same level of security effort as that in the normal form game. 

Note that our approach to characterizing NE and SPE can be readily extended to models with facility-dependent cost parameters and less than perfect defense. We conjecture that a different set of techniques will be required  to tackle the more general situation in which the attacker can target multiple facilities at the same time; see \cite{dahan2015network} for related work in this direction. However, even when the attacker targets multiple facilities, one can find game parameters for which the defender is always strictly better off in the sequential game. 

Finally, we provide a brief discussion on rational learning dynamics, aimed at understanding how the outcome of the attacker-defender interaction -- which may or may not result in compromise of a facility (state) -- effects the ability of system users to learn about the realized state through a repeated use of the system. A key issue is that the uncertainty about the realized state can significantly impact the ability of users to make decisions to ensure that their long-term cost corresponds to the true usage cost of the system. We explain this issue using a simple  transportation network as an example, in which rational travelers (users) need to learn about the identity of the facility that is likely to be compromised using imperfect information about the attack and noisy realizations of travel time in each stage of a repeated routing game played over the network.

The results reported in this article contribute to the study on the allocation of defense resources on facilities against strategic adversaries, as discussed in \cite{powell2007defending} and \cite{bier2007choosing}. The underlying assumption that drives our analysis is that an attack on each facility can be treated independently for the purpose of evaluating its impact on the overall usage cost. Other papers that also make this assumption include \cite{bell2008attacker}, \cite{bier2013defending}, \cite{alderson2011solving}, and \cite{brown2006defending}. Indeed, when the impact of facility compromises are related to the network structure, facilities can no longer be treated independently, and the network structure becomes a crucial factor in analyzing the defense strategy (\cite{dziubinski2017you}). Additionally, network connections can also introduce the possibility of cascading failure among facilities, which is addressed in \cite{acemoglu2016network}, and \cite{goyal2014attack}. These settings are not covered by our model.

The paper is structured as follows: In Sec. \ref{Sec:Model}, we introduce the model of both games, and discuss the modeling assumptions. We provide preliminary results to facilitate our analysis in Sec. \ref{Sec:attack-defend}. Sec. \ref{sec:generic_case} characterizes NE, and  Sec. \ref{sequential_section} characterizes SPE. Sec. \ref{outcomes} compares both games. We discuss some extensions of our model and briefly introduce dynamic aspects in Sec. \ref{example_sec}. 

All proofs are included in the appendix.

\section{The Model} \label{Sec:Model}
\subsection{Attacker-Defender Interaction: Normal Form versus Sequential Games}\label{model_present}
Consider an infrastructure system modeled as a set of components (facilities) $\E$. To defend the system against an external malicious attack, the system operator (defender) can secure one or more facilities in $\E$ by investing in appropriate security technology. The set of facilities in question can include cyber or physical elements that are crucial to the functioning of the system. 
These facilities are potential targets for a malicious adversary whose goal is to compromise the overall functionality of the system by gaining unauthorized access to certain cyber-physical elements.  
The security technology can be a combination of proactive mechanisms (authentication and access control) or reactive ones (attack detection and response). Since our focus is on modeling the strategic interaction between the attacker and defender at a system level, we do not consider the specific functionalities of individual facilities or the protection mechanisms offered by various technologies. 

We now introduce our game theoretic model. Let us denote a pure strategy of the defender as $\sd \subseteq \E$, with $\sd \in \Sd = 2^{\E}$. The cost of securing any facility is given by the parameter $\cd \in \mathbb{R}_{>0}$. Thus, the total defense cost incurred in choosing a pure strategy $\sd$ is $|\sd| \cdot \cd$, where $|\sd|$ is the cardinality of $\sd$ (i.e., the number of secured facilities). The attacker chooses to target a single facility $\e \in \E$ or not to attack. We denote a pure strategy of the attacker as $\sa \in \Sa =\E\cup \{\emptyset\}$. The cost of an attack is given by the parameter $\ca\in \mathbb{R}_{>0}$, and it reflects the effort that attacker needs to spend in order to successfully targets a single facility and compromise its operation. 

We assume that prior to the attack, the usage cost of the system is $\Czero$. This cost represents the level of efficiency with which the defender is able to operate the system for its users. A higher usage cost reflects lower efficiency. If a facility $\e$ is targeted by the attacker but not secured by the defender, we consider that $\e$ is compromised and the usage cost of the system changes to $\Ce$. Therefore, given any pure strategy profile $(\sd, \sa)$, the usage cost after the  attacker-defender interaction, denoted $C(\sd, \sa)$, can be expressed as follows:
\begin{align}\label{Ceq}
C(\sd, \sa)=\left\{
\begin{array}{ll}
\Ce, &\quad \text{if $\sa=\e$, and $\sd \not \owns \e$, }\\
\Czero, &\quad \text{otherwise.}
\end{array}
\right.
\end{align}

To study the effect of timing of the attacker-defender interaction, prior literature on security games has studied both normal form game and sequential games (\cite{alpcan2010network}). We study both models in our setting. In the normal form game, denoted $\gamen$,  the defender and the attacker move simultaneously. On the other hand, in the sequential game, denoted $\games$, the defender moves in the first stage and the attacker moves in the second stage after observing the defender's strategy. We allow both players to use mixed strategies. In $\gamen$, we denote the defender's mixed strategy as $\sigdn \deleq \(\sigdn(\sd)\)_{\sd \in \Sd} \in \Delta(\Sd)$, where $\sigdn(\sd)$ is the probability that the set of secured facilities is $\sd$. Similarly, a mixed strategy of the attacker is $\sigan \deleq \(\sigan(\sa)\)_{\sa \in \Sa} \in \Delta(\Sa)$, where $\sigan(\sa)$ is the probability that the realized action is $\sa$. Let $\sigma=\(\sigdn, \sigan\)$ denote a mixed strategy profile. In $\games$, the defender's mixed strategy $\sigds \deleq \(\sigds(\sd)\)_{\sd \in \Sd} \in \Delta(\Sd)$ is defined analogously to that in $\gamen$. The attacker's strategy is a map from $\Delta(\Sd)$ to $\Delta(\Sa)$, denoted by $\sigas(\sigds) \deleq \(\sigas(\sa, \sigds)\)_{\sa \in \Sa} \in \Delta(\Sa)$, where $\sigas(\sa, \sigds)$ is the probability that the realized action is $\sa$ when the defender's strategy is $\sigds$. A strategy profile in this case is denoted as $\sigs = \(\sigds, \sigas(\sigds)\)$. 

The defender's utility is comprised of two parts: the negative of the usage cost as given in \eqref{Ceq} and the defense cost incurred in securing the system. Similarly, the attacker's utility is the usage cost net the attack cost. For a pure strategy profile $\(\sd, \sa\)$, the utilities of defender and attacker can be respectively expressed as follows: 
\begin{align*}
\ud(\sd, \sa)&=-\ceq(\sd, \sa)-\cd \cdot |\sd|, \quad \ua(\sd, \sa)=\ceq(\sd, \sa)-\ca \cdot \mathds{1}\{\sa \neq \emptyset\}.
\end{align*}

For a mixed strategy profile $\(\sigdn, \sigan\)$, the expected utilities can be written as:
\begin{subequations}\label{U_fun}
\begin{align}
\Ud(\sigdn, \sigan)&=\sum_
{\sd \in \Sd} \sum_{\sa \in \Sa}\ud(\sd, \sa)\cdot \sigan(\sa) \cdot \sigdn(\sd)=-\mathbb{E}_{\sign} [\ceq]-\cd \cdot\mathbb{E}_{\sigdn} [|\sd|],\label{Ud} \\
\Ua(\sigdn, \sigan)&=\sum_
{\sd \in \Sd} \sum_{\sa \in \Sa}\ua(\sd, \sa)\cdot \sigan(\sa) \cdot \sigdn(\sd) =\mathbb{E}_{\sign} [\ceq]-\ca \cdot \mathbb{E}_{\sigan} [|\sa|],\label{Ua} 
\end{align}
\end{subequations}
where $\mathbb{E}_{\sign} [\ceq]$ is the expected usage cost, and $\mathbb{E}_{\sigdn}[|\sd|]$ (resp. $\mathbb{E}_{\sigan} [|\sa|]$) is the expected number of defended (resp. targeted) facilities, i.e.:
\begin{align*}
\mathbb{E}_{\sign} [\ceq]&=\sum_
{\sa \in \Sa} \sum_{\sd \in \Sd}\ceq(\sd, \sa)\cdot \sigan(\sa) \cdot \sigdn(\sd), \\
\mathbb{E}_{\sigdn}[|\sd|] &=\sum_{\sd \in \Sd}|\sd| \sigdn(\sd), \quad \mathbb{E}_{\sigan} [|\sa|]=\sum_{\e \in \E}\sigan(\e).
\end{align*}

An equilibrium outcome of the game $\gamen$ is defined in the sense of Nash Equilibrium (NE). A strategy profile $\signwe=(\sigdnwe, \siganwe)$ is a NE if:
\begin{align*}
\Ud(\sigdnwe, \siganwe) &\geq \Ud(\sigdn, \siganwe), \quad \forall \sigdn \in \Delta(\Sd),\\
\Ua(\sigdnwe, \siganwe) &\geq \Ua(\sigdnwe, \sigan), \quad \forall \sigan \in \Delta (\Sa).
\end{align*}

In the sequential game $\games$, the solution concept is that of a Subgame Perfect Equilibrium (SPE), which is also known as Stackelberg equilibrium. A strategy profile $\sigswe=(\sigdswe, \sigaswe(\sigds))$ is a SPE if:
\begin{subequations}\label{SPE_utility}
\begin{align}
\Ud(\sigdswe, \sigaswe(\sigdswe)) &\geq \Ud(\sigds, \sigaswe(\sigds)), \quad \forall \sigds \in \Delta(\Sd),\label{SPE:def}\\
\Ua(\sigds, \sigaswe(\sigds)) &\geq \Ua(\sigds, \sigas(\sigds)), \quad \forall \sigds \in \Delta(\Sd), \quad \forall \sigas(\sigds) \in \Delta(\Sa). \label{SPE:subgame}
\end{align}
\end{subequations}
Since both $\Sd$ and $\Sa$ are finite sets, and we consider mixed strategies, both NE and SPE exist. 

\subsection{Model Discussion}\label{assumption}
One of our main assumptions is that the attacker's capability is limited to targeting at most one facility, while the defender can invest in securing multiple facilities. Although this assumption appears to be somewhat restrictive, it enables us to derive analytical results on the equilibrium structure for a system with multiple facilities. The assumption can be justified in situations where the attacker can only target system components in a localized manner. Thus, a facility can be viewed as a set of collocated components that can be simultaneously targeted by the attacker. For example, in a transportation system, a facility can be a vulnerable link (edge), or a set of links that are connected by a vulnerable node (an intersection or a hub). In Sec. \ref{extension}, we briefly discuss the issues in solving a more general game where multiple facilities can be simultaneously targeted by the attacker.

Secondly, our model assumes that the costs of attack and defense are identical across all facilities. We make this assumption largely to avoid the notational burden of analyzing the effect of facility-dependent attack/defense cost parameters on the equilibrium structures. In fact, as argued in Sec. \ref{extension}, the qualitative properties of equilibria still hold when cost parameters are facility-dependent. However, characterizing the equilibrium regimes in this case can be quite tedious, and may not necessarily provide new insights on strategic defense investments. 

Thirdly, we allow both players to choose mixed strategies. Indeed, mixed strategies are commonly considered in security games as a pure NE may not always exists. A mixed strategy entails a player's decision to introduce randomness in her behavior, i.e. the manner in which a facility is targeted (resp. secured) by the attacker (resp. defender). Consider for example, the problem of inspecting a transportation network facing risk of a malicious attack. In this problem, a mixed strategy can be viewed as randomized allocation of inspection effort on subsets of facilities. Mixed strategy of the attacker can be similarly interpreted.

Fourthly, we assume that the defender has the technological means to perfectly secure a facility. In other words, an attack on a secured facility cannot impact its operation. As we will see in Sec. \ref{Sec:attack-defend}, the defender's mixed strategy can be viewed as the level of security effort on each facility, where the effort level 1 (maximum) means perfect security, and 0 (minumum) means no security. Under this interpretation, the defense cost $\cd$ is the cost of perfectly securing a unit facility (i.e., with maximum level of effort), and the expected defense cost is $\cd$ scaled by the security effort defined by the defender's mixed strategy.

Fifthly, we do not consider a specific functional form for modeling the usage cost. In our model, for any facility $\e\in \E$, the difference between the post-attack usage cost $\Ce$ and the pre-attack cost $\Czero$ represents the change of the usage cost of the system when $\e$ is compromised. This change can be evaluated based on the type of attacker-defender interaction one is interested in studying. For example, in situations when attack on a facility results in its complete disruption, one can use a connectivity-based metric such as the number of active source-destination paths or the number of connected components to evaluate the usage cost (\cite{dziubinski2013network} and \cite{dziubinski2017you}). On the other hand, in situations when facilities are congestible resources and an attack on a facility increases the users' cost of accessing it, the system's usage cost can be defined as the average cost for accessing (or routing through) the system. This cost can be naturally evaluated as the user cost in a Wardrop equilibrium (\cite{bier2013defending}), although socially optimal cost has also been considered in the literature (\cite{alderson2017assessing}).

Finally, we note that for the purpose of our analysis, the usage cost as given in \eqref{Ceq} fully captures the impact of player' actions on the system. For any two facilities $\e, \ep \in \E$, the ordering of $\Ce$ and $C_{\ep}$ determines the relative scale of impact of the two facilities. As we show in Sec. \ref{sec:generic_case}--\ref{sequential_section}, the order of cost functions in the set $\{C_e\}_{e\in\mathcal E}$ plays a key role in our analysis approach. Indeed the usage cost is intimately linked with the network structure and way of operation (for example, how individual users are routed through the network and how their costs are affected by a compromised facility). Barring a simple (yet illustrative) example, we do not elaborate further on how the network structure and/or the functional form of usage cost changes the interpretations of equilibrium outcome. We also do not discuss the computational aspects of arriving at the ordering of usage costs.

\section{Rationalizable Strategies and Aggregate Defense Effort}\label{Sec:attack-defend}
We introduce two preliminary results that are useful in our subsequent analysis. Firstly, we show that the defender's strategy can be equivalently represented by a vector of facility-specific security effort levels. Secondly, we identify the set of rationalizable strategies of both players. 

For any defender's mixed strategy $\sigdn \in \Delta(\Sd)$, the corresponding \emph{security effort vector} is $\pd=\(\ped\)_{\e \in \E}$, where $\ped$ is the probability that facility $\e$ is secured: 
\begin{align}\label{eq:ped}
\ped=\sum_{\sd \ni \e} \sigdn(\sd). 
\end{align}
In other words, $\ped$ is the level of security effort exerted by the defender on facility $\e$ under the security plan $\sigdn$. Since $\sigdn(\sd) \geq 0$ for any $\sd \in \Sd$, we obtain that $0 \leq \ped=\sum_{\sd \ni \e} \sigdn(\sd) \leq \sum_{\sd \in \Sd} \sigdn(\sd) = 1$. Hence, any $\sigdn$ induces a valid probability vector $\rho \in [0,1]^{|\E|}$. In fact, any vector $\rho \in [0, 1]^{|\E|}$ can be induced by at least one feasible $\sigdn$. The following lemma provides a way to explicitly construct one such feasible strategy. 
\begin{lemma}\label{lemma:stra_construct}
Consider any feasible security effort vector $\rho \in [0, 1]^{|\E|}$. Let $m$ be the number of distinct positive values in $\rho$, and define $\porderi$ as the $i$-th largest distinct value in $\rho$, i.e. $\porderone > \dots > \porderm$. The following defender's strategy is feasible and induces $\rho$:
\begin{subequations}\label{stra_construct}
\begin{align}
&\sigdn(\left\{\e \in \E| \pee \geq \porderi\right\})=\porderi-\porderipone, \quad \forall i=1, \dots, m-1\\
&\sigdn(\left\{\e \in \E| \pee \geq \porderm \right\})= \porderm, \\
&\sigdn(\emptyset)=1-\porderone.
\end{align}
For any remaining $\sd \in \Sd$, $\sigdn(\sd)=0$.  
\end{subequations}
\end{lemma}

We now re-express the player utilities in \eqref{U_fun} in terms of $\(\pd, \sigan\)$ as follows:
\begin{subequations}\label{U_fun_rewrite}
\begin{align}
\Ud(\sigdn, \sigan)&=-\sum_{\sa \in \Sa} \(\sum_
{\sd \in \Sd} \sigdn(\sd) C(\sd, \sa)\)\sigan(\sa)-\(\sum_{\sd \in \Sd}|\sd|\sigdn(\sd)\) \cd \notag \\
&=-\sum_{\e \in \E} \(\sum_
{\sd \in \Sd} \sigdn(\sd) C(\sd, \e)\)\sigan(\e)- \Czero \sigan(\emptyset)-\(\sum_{e \in \E} \ped\) \cd\notag \\
&\stackrel{\eqref{Ceq}}{=}-\sum_{\e \in \E} \(\(\sum_{\sd \ni \e}\sigdn(\sd)\) \Czero+\(1-\sum_{\sd \ni \e}\sigdn(\sd)\) C_e\)\sigan(\e)-\Czero \sigan(\emptyset) \notag\\
& \quad ~ -\(\sum_{e \in \E} \ped\) \cd \notag\\
&=-\sum_{\e \in \E} \(\ped \(\(\Czero-\Ce\) \sigan(\e) +\cd\) +\Ce \sigan(\e)\)-\Czero \sigan(\emptyset), \label{Ud_rewrite}\\
\Ua(\sigdn, \sigan)&=\sum_{\e \in \E} \(\ped \(\Czero-\Ce\) \sigan(\e) +\Ce \sigan(\e)\)+\Czero \sigan(\emptyset) - \(\sum_{\e \in \E} \sigan(\e)\)\ca. \label{Ua_rewrite}
\end{align}
\end{subequations}
Thus, for any given attack strategy and any two defense strategies, if the induced security effort vectors are identical, then the corresponding utility for each player is also identical. Henceforth, we denote the player utilities as $\Ud(\rho, \sigan)$ and $\Ua(\rho, \sigan)$, and use $\sigdn$ and $\ped$ interchangeably in representing the defender's strategy. 
For the sequential game $\games$, we analogously denote the security effort vector given the strategy $\sigds$ as $\psss(\sigds)$, and the defender's utility (resp. attacker's utility) as $\Uds(\psss, \sigas)$ (resp. $\Uas(\psss, \sigas)$).



We next characterize the set of rationalizable strategies. Note that the post-attack usage cost $\Ce$ can increase or remain the same or even decrease, in comparison to the pre-attack cost $\Czero$. Let the facilities whose damage result in an increased usage cost be grouped in the set $\Ebar$. Similarly, let $\Ehat$ denote the set of facilities such that a damage to any one of them has no effect on the usage cost. Finally, the set of remaining facilities is denoted as $\Edag$. Thus:      
\begin{subequations}
\begin{align}
\Ebar &\deleq \left\{\e \in \E | \Ce>\Czero\right\}, \label{Ebar}\\
\Ehat&\deleq \left\{\e \in \E \vert \Ce=\Czero\right\},\label{Ehat}\\
\Edag &\deleq \left\{\e \in \E \vert \Ce<\Czero\right\}. \label{Edag}
\end{align}
\end{subequations}
Clearly, $\Ebar \cup \Ehat \cup \Edag=\E$. The following proposition shows that in a rationalizable strategy profile, the defender does not secure facilities that are not in $\Ebar$, and the attacker only considers targeting the facilities that are in $\Ebar$. 
\begin{proposition}\label{strict_dominated}
The rationalizable action sets for the defender and attacker are given by $2^{\Ebar}$ and $\Ebar \cup \{\emptyset\}$, respectively. Hence, any equilibrium strategy profile $\(\pwe, \siganwe\)$ in $\gamen$ (resp. $\(\pswe, \sigaswe\)$ in $\games$) satisfies:
\begin{alignat*}{2}
\pewe&=\siganwe(\e)=0, &&\quad \forall \e \in \E \setminus \Ebar, \\
\psewe&=\sigaswe(\e, \psss)=0, &&\quad \forall \e \in \E \setminus \Ebar, \quad \forall \psss \in [0, 1]^{\E}.
\end{alignat*}
\end{proposition}

If $\Ebar=\emptyset$, then the attacker/defender does not attack/secure any facility in equilibrium. Henceforth, to avoid triviality, we assume $\Ebar \neq \emptyset$. Additionally, we define a partition of facilities in $\Ebar$ such that all facilities with identical $C_e$ are grouped in the same set. Let the number of distinct values in $\{\Ce\}_{\e \in \Ebar}$ be $\Ebarp$, and $\Cpi$ denote the $k$-th highest distinct value in the set $\{\Ce\}_{\e \in \Ebar}$. Then, we can order the usage costs as follows: 
\begin{align}\label{order}
C_{(1)} > C_{(2)} > \dots > C_{(\Ebarp)}>\Czero.
\end{align}
We denote $\Ebari$ as the set of facilities such that if any $\e \in \Ebari$ is damaged, the usage cost $\Ce=\Cpi$, i.e. $\Ebari \deleq \left\{\e \in \Ebar| \Ce=\Cpi\right\}$. We also define $E_{(k)} \deleq |\Ebari|$. Clearly, $\cup_{k=1}^{\Ebarp} \Ebari=\Ebar$, and $\sum_{k=1}^{\Ebarp} E_{(k)}=|\Ebar|$. Facilities in the same group have identical impact on the infrastructure system when compromised.

\section{Normal Form Game $\gamen$}\label{sec:generic_case}
In this section, we provide complete characterization of the set of NE for any given attack and defense cost parameters in game $\gamen$. In Sec. \ref{zero_sum_gamen}, we show that $\gamen$ is strategically equivalent to a zero-sum game, and hence the set of attacker's equilibrium strategies can be solved by a linear program. In Sec. \ref{in_regime}, we show that the space of cost parameters $(\ca, \cd) \in \mathbb{R}_{>0}^2$ can be partitioned into qualitatively distinct equilibrium regimes.
 \subsection{Strategic Equivalence to Zero-Sum Game}\label{zero_sum_gamen} 
Our notion of strategic equivalence is the same as the best-response equivalence defined in \cite{rosenthal1974correlated}. If $\gamen$ and another game $\gamezero$ are strategically equivalent, then given any strategy of the defender (resp. attacker), the set of attacker's (resp. defender's) best responses is identical in the two games. This result forms the basis of characterizing the set of NE.

We define the utility functions of the game $\gamezero$ as follows:
\begin{subequations}\label{zero_utility}
\begin{align}
\Udzero(\sigdn, \sigan)&=-\mathbb{E}_{\sign} [\ceq]-\mathbb{E}_{\sigdn}[|\sd|] \cdot \cd+ \ca \cdot \mathbb{E}_{\sigan}[|\sa|],\label{zero_utility_defend}\\
\Uazero(\sigdn, \sigan)&=\mathbb{E}_{\sign} [\ceq]+\mathbb{E}_{\sigdn}[|\sd|]\cdot \cd-\ca \cdot \mathbb{E}_{\sigan}[|\sa|]. \label{zero_utility_attack}
\end{align}
\end{subequations}
Thus, $\gamezero$ is a zero-sum game. We denote the set of defender's (resp. attacker's) equilibrium strategies in $\gamezero$ as $\Sigma_d^0$ (resp. $\Sigma_a^0$). 
\begin{lemma}\label{zero_sum}
The normal form game $\gamen$ is strategically equivalent to the zero sum game $\gamezero$. The set of defender's (resp. attacker's) equilibrium strategies in $\gamen$ is $\Sigma^{*}_d \equiv \Sigma_d^{0}$ (resp. $\Siganwe \equiv \Sigma_a^{0}$). Furthermore, for any $\sigdnwe \in \Sigma^{*}_d$ and any $\siganwe \in \Sigma^{*}_a$, $(\sigdnwe, \siganwe)$ is an equilibrium strategy profile of $\gamen$. 
\end{lemma}

Based on Lemma \ref{zero_sum}, the set of attacker's equilibrium strategies $\Siganwe$ can be expressed as the optimal solution set of a linear program.

\begin{proposition}\label{opt_eq}
The set $\Siganwe$ is the optimal solution set of the following optimization problem:
\begin{subequations}\label{maxmin_min}
\begin{align}
\max_{\sigan} \quad &V(\sigan) \notag \\
s.t. \quad & V(\sigan)=\sum_{\e \in \Ebar} \min \left\{\sigan(\e)\cdot \( \Czero-\ca\)+\cd,~ \sigan(\e) \cdot \(\Ce-\ca\)\right\}+\sigan(\emptyset) \cdot \Czero, \label{V}\\
&\sum_{\e \in \Ebar} \sigan(\e) +\sigan(\emptyset)=1, \label{feasible_one}\\
& \sigan(\emptyset) \geq 0, \quad \sigan(\e) \geq 0, \quad \forall \e \in \Ebar. \label{feasible_two}
\end{align}
\end{subequations}
Furthermore, \eqref{maxmin_min} is equivalent to the following linear optimization program:
\begin{subequations}\label{linear_maxmin}
\begin{align}
\max_{\sigan, v} \quad &\sum_{\e \in \Ebar} \ve+\sigan(\emptyset) \cdot \Czero \notag\\
s.t. \quad & \sigan(\e) \cdot \(\Czero-\ca\)+\cd-\ve \geq 0, \quad \forall \e \in \Ebar, \label{bound_1}\\
& \sigan(\e) \cdot \(\Ce-\ca\)-\ve \geq 0, \quad \forall \e \in \Ebar,\label{bound_2}\\
&\sum_{\e \in \Ebar} \sigan(\e)+\sigan(\emptyset)=1,\label{sum_signa}\\
& \sigan(\emptyset) \geq 0, \quad \sigan(\e) \geq 0, \quad \forall \e \in \Ebar. \label{non-negative}
\end{align}
\end{subequations}
where $v=\(\ve\)_{\e \in \Ebar}$ is an $|\Ebar|$-dimensional variable. 
\end{proposition}


In Proposition \ref{opt_eq}, the objective function $V(\sigan)$ is a piecewise linear function in $\sigan$. Furthermore, given any $\sigan$ and any $\e \in \Ebar$, we can write:
\begin{align}
&\min \left\{\sigan(\e) \cdot \(\Czero-\ca\)+\cd,~ \sigan(\e) \cdot \(\Ce-\ca\)\right\}\notag\\
=&
\left\{
\begin{array}{ll}
\sigan(\e) \cdot \(\Czero-\ca\)+\cd & \quad \text{if $\sigan(\e) > \frac{\cd}{\Ce-\Czero}$,}\\
\sigan(\e)\cdot \(\Ce-\ca\) & \quad \text{if $\sigan(\e) \leq \frac{\cd}{\Ce-\Czero}$.}
\end{array}
\right.\label{threshold}
\end{align}
Thus, we can observe that if $\sigan(\e)$ equals to $\cd/\(\Ce-\Czero\)$, then $-\sigan(\e) \cdot \Czero-\cd= -\sigan(\e) \cdot \Ce$, i.e. if a facility $\e$ is targeted with the threshold attack probability $\cd/(\Ce-\Czero)$, the defender is indifferent between securing $\e$ versus not.  The following lemma analyzes the defender's best response to the attacker's strategy, and shows that no facility is targeted with probability higher than the threshold probability in equilibrium. 
\begin{lemma}\label{only_attacked}
Given any strategy of the attacker $\sigan \in \Delta(\Sa)$, for any defender's security effort $\rho$ that is a best response to $\sigan$, denoted $\rho \in BR(\sigan)$, the security effort $\rho_e$ on each facility $\e \in \E$ satisfies:
\begin{align}\label{best_response_normal}
\rho_\e \left\{
\begin{array}{ll}
=0, &\quad \forall \e \in \left\{\Ebar|\sigan(\e)<\frac{\cd}{\Ce-\Czero}\right\} \cup \Ehat \cup \Edag, \\
\in [0, 1], & \quad \forall \e \in \left\{\Ebar|\sigan(\e)=\frac{\cd}{\Ce-\Czero}\right\}, \\
=1, &\quad \forall \e \in \left\{\Ebar|\sigan(\e)>\frac{\cd}{\Ce-\Czero}\right\}.
\end{array}
\right.
\end{align}
Furthermore, in equilibrium, the attacker's strategy $\siganwe$ satisfies:
\begin{subequations}\label{upper_bound}
\begin{alignat}{2}
\siganwe(\e) &\leq \frac{\cd}{\Ce-\Czero}, &&\quad \forall \e \in \Ebar, 
\label{sub:upper_bound}\\
\siganwe(\e) & = 0, &&\quad \forall \e \in \E \setminus \Ebar. \label{zero_out}
\end{alignat}
\end{subequations}
\end{lemma}

Lemma \ref{only_attacked} highlights a key property of NE: The attacker does not target at any facility $\e\in \Ebar$ with probability higher than the threshold $\cd/(\Ce-\Czero)$, and the defender allocates a non-zero security effort only on the facilities that are targeted with the threshold probability.

Intuitively, if a facility $\e$ were to be targeted with a probability higher than the threshold $\cd/(\Ce-\Czero)$, then the defender's best response would be to secure that facility with probability 1, and the attacker's expected utility will be $-\Czero-\ca\sigan(\e)$, which is smaller than $-\Czero$ (utility of no attack). Hence, the attacker would be better off by choosing the no attack action.



Now, we can re-write $V(\sigan)$ as defined in \eqref{maxmin_min} as follows:
\begin{align}
V(\sigan)&\stackrel{\eqref{upper_bound}}{=}\sum_{\e \in \{\Ebar|\sigan(\e) \leq \frac{\cd}{\Ce-\Czero}\}} \sigan(\e) \(\Ce-\ca\)+\Czero \cdot \sigan(\emptyset),\label{re-express-V}
\end{align} 
and the set of attacker's equilibrium strategies maximizes this function. 
\subsection{Characterization of NE in $\gamen$}\label{in_regime}
We are now in the position to introduce the equilibrium regimes. Each regime corresponds to a range of cost parameters such that the qualitative properties of equilibrium (i.e. the set of facilities that are targeted and secured) do not change in the interior of each regime. 

We say that a facility $e$ is \emph{vulnerable} if $\Ce-\ca>\Czero$. Therefore, given any attack cost $\ca$, the set of vulnerable facilities is given by $\{\Ebar| \Ce-\ca>\Czero\}$. Clearly, only vulnerable facilities are likely targets of the attacker. If $\ca> C_{(1)}-\Czero$, then there are no vulnerable facilities. In contrast, if $\ca < C_{(1)}-\Czero$, we define the following threshold for the per-facility defense cost:
\begin{align}\label{cdbar}
\cdbar \deleq \frac{1}{\sum_{\e \in \{\Ebar| \Ce-\ca >\Czero\}} \frac{1}{\Ce-\Czero}}.
\end{align}
We can check that for any $i=1, \dots, \Ebarp-1$ (resp. $i=\Ebarp$), if $C_{(i+1)}-\Czero\leq\ca< C_{(i)}-\Czero$ (resp. $0<\ca<C_{(K)}-\Czero$), then 
\begin{align}\label{cd_accurate}
\cdbar=\(\sum_{k=1}^{i} \frac{\nnk}{C_{(k)}-\Czero}\)^{-1}.
\end{align}
Recall from Lemma \ref{only_attacked} that $\siganwe(\e)$ is upper bounded by the threshold attack probability $\cd/(\Ce-\Czero)$. If the defense cost $\cd<\cdbar$, then $\sum_{k=1}^{i} \frac{\nnk \cd}{C_{(k)}-\Czero}<1$, which implies that even when the attacker targets each vulnerable facility with the threshold attack probability, the total probability of attack is still smaller than 1. Thus, the attacker must necessarily choose not to attack with a positive probability. On the other hand, if $\cd>\cdbar$, then the no attack action is not chosen by the attacker in equilibrium. 

Following the above discussion, we introduce two types of regimes depending on whether or not $\cd$ is higher than the threshold $\cdbar$. In type I regimes, denoted as $\{\regimei | i=0, \dots, \Ebarp\}$, the defense cost $\cd<\cdbar$, whereas in type II regimes, denoted as $\{\regimej | j=1, \dots, \Ebarp\}$, the defense cost $\cd>\cdbar$. Hence, we say that $\cd$ is ``relatively low'' (resp. ``relatively high'') in comparison to $\ca$ in type I regimes (resp. type II regimes). We formally define these $2K+1$ regimes as follows:
\begin{enumerate}[label=(\alph*)]
\item Type I regimes $\regimei$, $i=0, \dots, \Ebarp$:
\begin{itemize}
\item If $i=0$: 
\begin{align}\label{regimei_first}
\ca> C_{(1)}-\Czero, \text{ and }\cd>0
\end{align}
\item If $i=1, \dots, \Ebarp-1$:
\begin{align}\label{regimei_notlast}
C_{(i+1)}-\Czero < \ca < C_{(i)}-\Czero, \text{ and }  0<\cd< \(\sum_{k=1}^{i} \frac{\nnk}{\Cpk-\Czero}\)^{-1}
\end{align}
\item If $i=\Ebarp$: 
\begin{align}\label{regimei_last}
0 < \ca < C_{(\Ebarp)}-\Czero, \text{ and } 0<\cd<\(\sum_{k=1}^{\Ebarp} \frac{\nnk}{\Cpk-\Czero}\)^{-1}
\end{align}
\end{itemize}
\item Type II regimes, $\regimej$, $j=1, \dots, \Ebarp$:  
\begin{itemize}
\item If $j=1$:
\begin{align}\label{regime_j_1}
0<\ca< C_{(1)}-\Czero, \text{ and } \cd> \(\frac{E_{(1)}}{C_{(1)}-\Czero}\)^{-1}
\end{align}
\item If $j=2, \dots, \Ebarp$:
\begin{align}\label{regime_j_rest}
0<\ca< C_{(j)}-\Czero, \text{ and }  \(\sum_{k=1}^{j} \frac{\nnk}{\Cpk-\Czero}\)^{-1}< \cd <\(\sum_{k=1}^{j-1} \frac{\nnk}{\Cpk-\Czero}\)^{-1}
\end{align}
\end{itemize}
\end{enumerate}
We now characterize equilibrium strategy sets $\Sigdnwe$ and $\Siganwe$ in the interior of each regime.\footnote{For the sake of brevity, we omit the discussion of equilibrium strategies when cost parameters lie exactly on the regime boundary, although this case can be addressed using the approach developed in this article.}
\begin{theorem}\label{attacker_strategy}
The set of NE in each regime is as follows:
\begin{enumerate}[label=(\alph*)]
\item Type I regimes $\regimei$: 
\begin{itemize}
\item If $i=0$, 
\begin{subequations}
\begin{align}
\pewe&=0, \quad \forall \e \in \E \label{defender_regime_i_0}\\
\siganwe(\emptyset)&=1. \label{unique_zero}
\end{align}
\end{subequations}
\item If $i=1, \dots, \Ebarp$, 
\begin{subequations}\label{attack_n_1}
\begin{alignat}{2}
\pewe&=\frac{\Cpk-\ca-\Czero}{\Cpk-\Czero},&& \quad \forall \e \in \Ebar_{(k)},  \quad \forall k=1, \dots, i \label{regime_last_sub}\\
\pewe&=0, &&\quad \forall \e \in \E \setminus \(\cup_{k=1}^{i} \Ebar_{(k)}\)\label{regime_last_zero}\\
\siganwe(\e)&= \frac{\cd}{\Cpi-\Czero}, &&\quad \forall \e \in \Ebari, \quad \forall k=1, \dots, i \label{multiple_bound}\\
\siganwe(\emptyset)&=1-\sum_{\e \in \cup_{k=1}^{i} \Ebar_{(k)}} \siganwe(\e). &&\label{6d}
\end{alignat}
\end{subequations}
\end{itemize}
\item Type II regimes $\regimej$:
\begin{itemize}
\item $j=1$: 
\begin{subequations}\label{attack_one}
\begin{alignat}{2}
\pewe&=0, &&\quad \forall \e \in \E \label{regime_1_unique_p}\\
0  \leq \siganwe(\e) &\leq \frac{\cd}{C_{(1)}-\Czero} &&\quad \forall \e \in \Ebar_{(1)}, \label{sub:upper_one}\\
\sum_{\e \in \Ebar_{(1)}}\siganwe(\e)&=1. &&\label{sum_j_one}
\end{alignat}
\end{subequations}
\item $j=2, \dots, \Ebarp$: 
\begin{subequations}\label{regime_k_attack}
\begin{alignat}{2}
\pewe&=\frac{\Cpk-C_{(j)}}{\Cpk-\Czero}, \quad \forall \e \in \Ebar_{(k)}, &&\quad \forall k=1, \dots, j-1 \label{defend_k_p}\\
\pewe&=0, &&\quad \forall \e \in \E \setminus \(\cup_{k=1}^{j-1} \Ebar_{(k)}\) \\
\siganwe(\e) &= \frac{\cd}{C_{(k)}-\Czero}, && \quad \forall \e \in \Ebar_{(k)}, \quad \forall k=1, \dots, j-1\label{regime_k_1}\\
0 \leq \siganwe(\e) &\leq \frac{\cd}{C_{(j)}-\Czero}, &&\quad \forall \e \in \Ebar_{(j)}\label{regime_k_2}\\
\sum_{\e \in \Ebar_{(j)}}\siganwe(\e)&=1-\sum_{k=1}^{j-1}\frac{\cd \cdot \nnk}{C_{(k)}-\Czero}.&& \label{regime_k_3}
\end{alignat}
\end{subequations}
\end{itemize}
\end{enumerate}
\end{theorem}

Let us discuss the intuition behind the proof of Theorem \ref{attacker_strategy}. 

Recall from Proposition \ref{opt_eq} and Lemma \ref{only_attacked} that the set of attacker's equilibrium strategies $\Sigma_a^{*}$ is the set of feasible mixed strategies that maximizes $V(\sigan)$ in \eqref{re-express-V}, and the attacker never targets at any facility $\e \in \E$ with probability higher than the threshold $\cd/(\Ce-\Czero)$. Also recall that the costs $\{C_{(k)}\}_{k=1}^K$ are ordered according to \eqref{order}. Thus, in equilibrium, the attacker targets the facilities in $\Ebar_{(k)}$ with the threshold attack probability starting from $k=1$ and proceeding to $k=2, 3, \dots \Ebarp$ until either all the vulnerable facilities are targeted with the threshold attack probability (and no attack is chosen with remaining probability), or the total attack probability reaches 1. 

Again, from Lemma \ref{only_attacked}, we know that the defender secures the set of facilities that are targeted with the threshold attack probability with positive effort. The equilibrium level of security effort ensures that the attacker gets an identical utility in choosing any pure strategy in the support of $\siganwe$, and this utility is higher or equal to that of choosing any other pure strategy. 

The distinctions between the two regime types are summarized as follows: 
\begin{enumerate}
\item In type I regimes, the defense cost $\cd< \cdbar$. The defender secures all vulnerable facilities with a positive level of effort. The attacker targets at each vulnerable facility with the threshold attack probability, and the total probability of attack is less than 1.
\item In type II regimes, the defense cost $\cd>\cdbar$. The defender only secures a subset of targeted facilities with positive level of security effort.  The attacker chooses the facilities in decreasing order of $\Ce-\Czero$, and targets each of them with the threshold probability until the attack resource is exhausted, i.e. the total probability of attack is 1.
\end{enumerate}

\section{Sequential game $\games$}\label{sequential_section}
In this section, we characterize the set of SPE in the game $\games$ for any given attack and defense cost parameters. The sequential game $\games$ is no longer strategically equivalent to a zero-sum game. Hence, the proof technique we used for equilibrium characterization in game $\gamen$ does not work for the game $\games$. In Sec. \ref{spe_preparation}, we analyze the attacker's best response to the defender's security effort vector. We also identify a threshold level of security effort which determines whether or not the defender achieves full attack deterrence in equilibrium. In Sec. \ref{spe_subsec}, we present the equilibrium regimes which govern the qualitative properties of SPE.  

\subsection{Properties of SPE}\label{spe_preparation}
By definition of SPE, for any security effort vector $\psss \in [0, 1]^{|\E|}$ chosen by the defender in the first stage, the attacker's equilibrium strategy in the second stage is a best response to $\psss$, i.e. $\sigaswe(\psss)$ satisfies \eqref{SPE:subgame}. As we describe next, the properties of SPE crucially depend on a threshold security effort level defined as follows: 
\begin{align}\label{pebar}
\pebar \deleq \frac{\Ce-\ca-\Czero}{\Ce-\Czero}, \quad \forall \e \in \Ebar.
\end{align}
The following lemma presents the best response correspondence $BR(\psss)$ of the attacker: 
\begin{lemma}\label{best_response_sequential}
Given any $\psss \in [0, 1]^{|\E|}$, if $\psss$ satisfies $\pess\geq \pebar$, for all $\e \in \{\Ebar | \Ce-\ca>\Czero\}$, then $BR(\psss)=\Delta(\Erho \cup \{\emptyset\})$, where:
\begin{align}\label{Erho}
\Erho \deleq \left\{\Ebar \left\vert \Ce-\ca>\Czero, \quad \pess= \pebar\right.\right\}.
\end{align}
Otherwise, $BR(\psss)=\Delta(\Emax)$, where:
\begin{align}\label{Emax}
\Emax\deleq\underset{\e \in \{\Ebar | \Ce-\ca>\Czero\}}{\mathrm{argmax}} \left\{\pess \Czero+ (1-\pess)\Ce \right\}.
\end{align}
\end{lemma}
In words, if each vulnerable facility $\e$ is secured with an effort higher or equal to the threshold effort $\pebar$ in \eqref{pebar}, then the attacker's best response is to choose a mixed strategy with support comprised of all vulnerable facilities that are secured with the threshold level of effort (i.e., $\Erho$ as defined in \eqref{Erho}) and the no attack action. Otherwise, the support of attacker's strategy is comprised of all vulnerable facilities (pure actions) that maximize the expected usage cost (see \eqref{Emax}). In particular, no attack action is not chosen in attacker's best response.  

Now recall that any SPE $(\pswe, \sigaswe(\pswe))$ must satisfy both \eqref{SPE:def} and \eqref{SPE:subgame}. Thus, for an equilibrium security effort $\pswe$, an attacker's best response $\sigas(\pswe) \in BR(\pswe)$ is an equilibrium strategy only if both these constraints are satisfied. The next lemma shows that depending on whether the defender secures each vulnerable facility $\e$ with the threshold effort $\pebar$ or not, the total attack probability in equilibrium is either 0 or 1. Thus, the defender being the first mover determines
whether the attacker is fully deterred from conducting an attack or not. Additionally, in SPE, the security effort on each vulnerable facility $\e$ is no higher than the threshold effort $\pebar$, and the security effort on any other edge is 0. 

\begin{lemma}\label{zero_or_one}
Any SPE $(\pswe, \sigaswe(\pswe))$ of the game $\games$ satisfies the following property: 
\begin{align*} 
\sum_{\e \in \Ebar}\sigaswe(\e, \pswe)= \left\{
\begin{array}{ll}
0, & \quad \text{if $\psewe\geq  \pebar, \quad \forall \e \in \{\Ebar | \Ce-\ca>\Czero\}$},\\
1,  & \quad  \text{otherwise.}
\end{array}
\right.
\end{align*}
Additionally, for any $\e \in \{\Ebar| \Ce-\ca>\Czero\}$, $\psewe \leq \pebar$. For any $\e \in \E \setminus \{\Ebar| \Ce-\ca>\Czero\}$, $\psewe =0$.
\end{lemma}
The proof of this result is based on the analysis of following three cases: 

\noindent\underline{Case 1}: There exists at least one facility $\e \in \{\Ebar|\Ce-\ca>\Czero\}$ such that $\psewe<\pebar$. In this case, by applying Lemma \ref{best_response_sequential}, we know that $\sigaswe(\pswe) \in BR(\pswe) = \Delta(\Emax)$, where $\Emax$ is defined in \eqref{Emax}. Hence, the total attack probability is 1. 

\noindent\underline{Case 2}: For any $\e \in \{\Ebar|\Ce-\ca>\Czero\}$, $\psewe>\pebar$. In this case, the set $\Erho$ defined in \eqref{Erho} is empty. Hence, Lemma \ref{best_response_sequential} shows that the total attack probability is 0.

\noindent\underline{Case 3}: For any $\e \in \{\Ebar|\Ce-\ca>\Czero\}$, $\psewe \geq \pebar$, and the set $\Erho$ in \eqref{Erho} is non-empty. Again from Lemma \ref{best_response_sequential}, we know that $\sigaswe(\pswe) \in BR(\pswe) = \Delta(\Erho \cup \{\emptyset\})$. Now assume that the attacker chooses to target at least one facility $\e \in \Erho$ with a positive probability in equilibrium. Then, the defender can deviate by slightly increasing the security effort on each facility in $\Erho$. By introducing such a deviation, the defender's security effort satisfies the condition of Case 2, where the total attack probability is 0. Hence, this results in a higher utility for the defender. Therefore, in any SPE $\(\pswe, \sigaswe(\pswe)\)$, one cannot have a second stage outcome in which the attacker targets facilities in $\Erho$. We can thus conclude that the total attack probability must be 0 in this case. 

In both Cases 2 and 3, we say that the attacker is \emph{fully deterred}. 

Clearly, these three cases are exhaustive in that they cover all feasible security effort vectors, and hence we can conclude that the total attack probability in equilibrium is either 0 or 1. Additionally, since the attacker is fully deterred when each vulnerable facility is secured with the threshold effort, the defender will not further increase the security effort beyond the threshold effort on any vulnerable facility. That is, only Cases 1 and 3 are possible in equilibrium. 

\subsection{Characterization of SPE}\label{spe_subsec}
Recall that in Sec. \ref{sec:generic_case}, type I and type II regimes for the game $\gamen$ can be distinguished based on a threshold defense cost $\cdbar$. It turns out that in $\games$, there are still $2 \Ebarp+1$ regimes. Again, each regime denotes distinct ranges of cost parameters, and can be categorized either as type $\typeone$ or type $\typetwo$. However, in contrast to $\gamen$, the regime boundaries in this case are more complicated; in particular, they are non-linear in the cost parameters $\ca$ and $\cd$.

To introduce the boundary $\cdtil(\ca)$, we need to define the function $\cdij(\ca)$ for each $i=1, \dots, K$ and $j=1, \dots, i$ as follows: 
\begin{align}\label{cdij}
\cdij(\ca) = \left\{
\begin{array}{ll}
\frac{C_{(1)}-\Czero}{\sum_{k=1}^{i} E_{(k)}-\sum_{k=1}^i\frac{\ca E_{(k)}}{\Cpk-\Czero}}, & \quad \text{if } j=1,\\
&\\
\frac{C_{(j)}-\Czero}{\(C_{(j)}-\Czero\) \cdot \(\sum_{k=1}^{j-1} \frac{E_{(k)}}{\Cpk-\Czero}\) + \sum_{k=j}^{i} E_{(k)}-\sum_{k=1}^{i} \frac{\ca E_{(k)}}{\Cpk-\Czero}}, & \quad \text{if } j=2, \dots, i.
\end{array}
\right.
\end{align}
For any $i=1, \dots, K$, and any attack cost $C_{(i+1)}-\Czero \leq \ca< C_{(i)}-\Czero$, but $0<\ca<C_{(K)}-\Czero$ if $i=K$, the threshold $\cdtil(\ca)$ is defined as follows: 
\begin{align}\label{cdtil}
\cdtil(\ca)=\left\{
\begin{array}{ll}
\cdij(\ca), & \quad \text{if $\frac{\sum_{k=j+1}^{i}E_{(k)}}{\sum_{k=1}^{i} \frac{E_{(k)}}{\Cpk-\Czero}}\leq \ca<\frac{\sum_{k=j}^{i}E_{(k)}}{\sum_{k=1}^{i} \frac{E_{(k)}}{\Cpk-\Czero}}$, and $j=1, \dots, i-1$,}\\
\cd^{ii}(\ca), & \quad \text{if $0\leq \ca<\frac{E_{(i)}}{\sum_{k=1}^{i} \frac{E_{(k)}}{\Cpk-\Czero}}$}.
\end{array}
\right.
\end{align}
\begin{lemma}\label{comparison_lemma}
Given any attack cost $0\leq\ca<C_{(1)}-\Czero$, the threshold $\cdtil(\ca)$ is a strictly increasing and continuous function of $\ca$.

Furthermore, for any $0<\ca<C_{(1)}-\Czero$, $\cdtil(\ca)>\cdbar$. If $\ca=0$, $\cdtil(0)=\bar{\cd}(0)$. If $\ca \to C_{(1)}-\Czero$, $\cdtil(\ca) \to +\infty$. 
\end{lemma}
Since $\cdtil(\ca)$ is a strictly increasing and continuous function function of $\ca$, the inverse function $\cdtilinv(\cd)$ is well-defined. Now we are ready to formally define the regimes for the game $\games$:

\begin{enumerate}
\item Type $\typeone$ regimes $\regimesi$, $i=0, \dots, \Ebarp$:
\begin{itemize}
\item If $i=0$:
\begin{align}\label{regimesi_first}
\ca>C_{(1)}-\Czero, \text{ and } \quad \cd>0.
\end{align}
\item If $i=1, \dots, \Ebarp-1$: 
\begin{align}\label{regimesi_middle}
C_{(i+1)}-\Czero <\ca< C_{(i)}-\Czero, \text{ and } \quad 0<\cd< \cdtil(\ca).
\end{align}
\item If $i=\Ebarp$: 
\begin{align}\label{regimesi_last}
0<\ca< C_{(\Ebarp)}-\Czero, \text{ and } \quad  0<\cd< \cdtil(\ca).
\end{align}
\end{itemize}
\item Type $\typetwo$ regimes $\regimesj$, $j=1, \dots, \Ebarp$:
\begin{itemize}
\item If $j=1$:
\begin{align}\label{regimej_constraint_1}
0< \ca< \cdtilinv(\cd),  \text{ and } \quad \cd> \(\frac{E_{(1)}}{C_{(1)}-\Czero}\)^{-1}
\end{align}
\item If $j=2, \dots, \Ebarp$: 
\begin{align}\label{regimej_constraint}
0< \ca< \cdtilinv(\cd), \text{ and } \quad \(\sum_{k=1}^{j} \frac{E_{(k)}}{\Cpk-\Czero}\)^{-1}<\cd<\(\sum_{k=1}^{j-1} \frac{E_{(k)}}{\Cpk-\Czero}\)^{-1}
\end{align}
\end{itemize}
\end{enumerate}

Analogous to the discussion in Section \ref{in_regime}, we say $\cd$ is ``relatively low'' in type $\typeone$ regimes, and ``relatively high'' in type $\typetwo$ regimes. We now provide full characterization of SPE in each regime. 
\begin{theorem}\label{theorem:SPE}
The defender's equilibrium security effort vector $\pswe=\(\psewe\)_{\e \in \E}$ is unique in each regime. Specifically, SPE in each regime is as follows:
\begin{enumerate}
\item Type $\typeone$ regimes $\regimesi$:
\begin{itemize}
\item If $i=0$, 
\begin{subequations}\label{SPE_i_0}
\begin{alignat}{2}
\psewe&=0, &&\quad \forall \e \in \E, \\
\sigaswe(\emptyset, \psss)&=1, &&\quad \forall \psss \in [0, 1]^{|\E|}.
\end{alignat}
\end{subequations}
\item If $i=1, \dots, \Ebarp$, 
\begin{subequations}\label{SPE_i}
\begin{alignat}{2}
\psewe&=\frac{\Cpk-\ca-\Czero}{\Cpk-\Czero}, && \quad \forall \e \in \Ebari, \quad \forall k=1, \dots, i,\label{SPE_i_defender_positive} \\
\psewe&=0, &&\quad \forall \e \in \E\setminus \(\cup_{k=1}^{i} \Ebari\),\label{SPE_i_defender_zero}\\
\sigaswe(\emptyset, \pswe)&=1,&&\\
\sigaswe(\psss) &\in BR(\psss), &&\quad \forall \psss \in [0, 1]^{|\E|}\setminus \pswe. \label{SPE_i_attacker}
\end{alignat}
\end{subequations}
\end{itemize}
\item Type $\typetwo$ regimes $\regimesj$:
\begin{itemize}
\item If $j=1$, 
\begin{subequations}\label{SPE_j_1}
\begin{alignat}{2}
\psewe&=0, &&\quad \forall \e \in \E,\\
\sigaswe(\pswe)&\in\Delta(\Ebar_{(1)}), &&\\
\sigaswe(\psss)&\in BR(\psss),&& \quad \forall \psss \in [0, 1]^{|\E|}\setminus \pswe. \label{SPE_j_1_attacker}
\end{alignat}
\end{subequations}
\item If $j=2, \dots, \Ebarp$, 
\begin{subequations}\label{SPE_j}
\begin{alignat}{2}
\psewe&=\frac{\Cpk-C_{(j)}}{\Cpk-\Czero}, &&\quad \forall \e \in \Ebari,\quad \forall k=1, \dots, j-1,\\
\psewe&=0, &&\quad \forall \e \in \E \setminus \(\cup_{k=1}^{j-1} \Ebari\),\\
\sigaswe(\pswe)&\in\Delta\(\cup_{k=1}^{j}\Ebar_{(k)}\), &&\\
\sigaswe(\psss)&\in BR(\psss), &&\quad \forall \psss \in [0, 1]^{|\E|}\setminus \pswe.\label{SPE_j_attacker}
\end{alignat}
\end{subequations}
\end{itemize}
\end{enumerate}
\end{theorem}
In our proof of Theorem \ref{theorem:SPE} (see Appendix \ref{proof_sequential}), we take the approach by first constructing a partition of the space $(\ca, \cd) \in \mathbb{R}_{>0}^2$ defined in \eqref{partition}, and then characterizing the SPE for cost parameters in each set in the partition (Lemmas \ref{sequential_type1}--\ref{type_2_sequential}). Theorem \ref{theorem:SPE} follows directly by regrouping/combining the elements of this partition such that each of the new partition has qualitatively identical equilibrium strategies.

From the discussion of Lemma \ref{zero_or_one}, we know that only Cases 1 and 3 are possible in equilibrium, and that in any SPE, the security effort on each vulnerable facility $\e$ is no higher than the threshold effort $\pebar$. It turns out that for any attack cost, depending on whether the defense cost is lower or higher than the threshold cost $\cdtil(\ca)$, the defender either secures each vulnerable facility with the threshold effort given by \eqref{cdtil} (type $\typeone$ regime), or there is at least one vulnerable facility that is secured with effort strictly less than the threshold (type $\typetwo$ regimes):

\begin{itemize}
\item In type $\typeone$ regimes, the defense cost $\cd<\cdtil(\ca)$. The defender secures each vulnerable facility with the threshold effort $\pebar$. The attacker is fully deterred. 
\item In type $\typetwo$ regimes, the defense cost $\cd>\cdtil(\ca)$. The defender's equilibrium security effort is identical to that in NE of the normal form game $\gamen$. The total attack probability is 1. 
\end{itemize}

\section{Comparison of $\gamen$ and $\games$}\label{outcomes}
Sec. \ref{utility_comparison} deals with the comparison of players' equilibrium utilities in the two games. In Sec. \ref{NE_SPE}, we compare the equilibrium regimes and discuss the distinctions in equilibrium properties of the two games. This leads us to an understanding of the effect of timing of play, i.e. we can identify situations in which the defender gains by proactively investing in securing all of the vulnerable facilities at an appropriate level of effort.

\subsection{Comparison of Equilibrium Utilities}\label{utility_comparison}
The equilibrium utilities in both games are unique, and can be directly derived using Theorems \ref{attacker_strategy} and \ref{theorem:SPE}. We denote the equilibrium utilities of the defender and attacker in regime $\regimei$ (resp. $\regimej$) as $\Ud^{\regimei}$ and $\Ua^{\regimei}$ (resp. $\Ud^{\regimej}$ and $\Ua^{\regimej}$) in $\gamen$, and $\Uds^{\regimesi}$ and $\Uas^{\regimesi}$ (resp. $\Uds^{\regimesj}$ and $\Uas^{\regimesj}$) in regime $\regimesi$ (resp. $\regimesj$) in $\games$. 
\begin{proposition}\label{utility}
In both $\gamen$ and $\games$, the equilibrium utilities are unique in each regime. Specifically, 
\begin{enumerate}[label=(\alph*)]
\item Type I $(\typeone)$ regimes $\regimei$ $(\regimesi)$:
\begin{itemize}
\item If $i=0$:
\begin{align*}
\Ud^{\Lambda_0}&=\Uds^{\widetilde{\Lambda}^{0}}=-\Czero, \text{ and } \quad
\Ua^{\Lambda_0}= \Uas^{\widetilde{\Lambda}^0}=\Czero.\\
\end{align*}
\item If $i=1, \dots, \Ebarp$:
\begin{alignat*}{2}
\Udi&=-\Czero-\(\sum_{k=1}^{i}  \nnk\)\cd, \quad &&\text{ and } \quad \Uai=\Czero, \\
\Uds^{\regimesi}&=-\Czero-\(\sum_{k=1}^{i} \frac{\(\Ce-\ca-\Czero\)\nnk}{\Ce-\Czero}\) \cd, \quad &&\text{and } \quad \Uas^{\regimesi}=\Czero.
\end{alignat*}
\end{itemize}
\item Type II ($\typetwo$) regimes $\regimej$ ($\regimesj$):
\begin{itemize}
\item If $j=1$:
\begin{align*}
\Ud^{\Lambda_1}&=\Uds^{\widetilde{\Lambda}_{1}}=-C_{(1)},\text{ and } \quad \Ua^{\Lambda_1}=\Uas^{\widetilde{\Lambda}_{1}}=C_{(1)}-\ca. 
\end{align*}
\item If $j=2, \dots, \Ebarp$: 
\begin{align*}
\Udj&=\Uds^{\regimesj}=-C_{(j)}-\sum_{k=1}^{j-1} \frac{\(\Cpk-C_{(j)}\) \cd\nnk}{\Cpk-\Czero} , \text{ and } \quad  \Uaj=\Uas^{\regimesj}=C_{(j)}-\ca.\\
\end{align*}
\end{itemize}
\end{enumerate}
\end{proposition}

From our results so far, we can summarize the similarities between the equilibrium outcomes in $\gamen$ and $\games$. While most of these conclusions are fairly intuitive, the fact that they are common to both game-theoretic models suggests that the timing of defense investments do not play a role as far as these insights are concerned. 
Firstly, the support of both players equilibrium strategies tends to contain the facilities, whose compromise results in a high usage cost. The defender secures these facilities with a high level of effort in order to reduce the probability with which they are targeted by the attacker. 
Secondly, the attack and defense costs jointly determine the set of facilities that are targeted or secured in equilibrium. On one hand, the set of vulnerable facilities increases as the cost of attack decreases. On the other hand, when the cost of defense is sufficiently high, the attacker tends to conduct an attack with probability 1. However, as the defense cost decreases, the attacker randomizes the attack on a larger set of facilities. Consequently, the defender secures a larger set of facilities with positive effort, and when the cost of defense is sufficiently small, all vulnerable facilities are secured by the defender. Thirdly, each player's equilibrium payoff is non-decreasing in the opponent's cost, and non-increasing in her own cost. Therefore, to increase her equilibrium payoff, each player is better off as her own cost decreases and the opponent's cost increases.

\subsection{First Mover Advantage}\label{NE_SPE}
We now focus on identifying parameter ranges in which the defender has the first mover advantage, i.e., the defender in SPE has a strictly higher payoff than in NE. To identify the first mover advantage, let us recall the expressions of type I regimes for $\gamen$ in \eqref{regimei_first}--\eqref{regimei_last} and type $\typeone$ regimes for $\games$ in \eqref{regimesi_first}--\eqref{regimesi_last}. Also recall that, for any given cost parameters $\ca$ and $\cd$, the threshold $\cdbar$ (resp. $\cdtil(\ca)$) determines whether the equilibrium outcome is of type I or type II regime (resp. type $\typeone$ or $\typetwo$ regime) in the game $\Gamma$ (resp. $\games$). Furthermore, from Lemma \ref{comparison_lemma}, we know that the cost threshold $\cdbar$ in $\gamen$ is smaller than the threshold $\cdtil(\ca)$ in $\games$. Thus, for all $i=1, \dots, \Ebarp$, the type I regime $\regimei$ in $\gamen$ is a proper subset of the type $\typeone$ regime $\regimesi$ in $\games$. Consequently, for any $\(\ca, \cd\) \in \mathbb{R}_{>0}^2$, we can have one of the following three cases:
\begin{itemize}
\item $0<\cd< \cdbar$: The defense cost is relatively low in both $\gamen$ and $\games$. We denote the set of $\(\ca, \cd\)$ that satisfy this condition as $\regimel$ (\emph{low} cost). That is, 
\begin{align}\label{L_set}
\regimel \deleq \left\{\(\ca, \cd\)|0<\cd< \cdbar \right\}=\cup_{i=0}^{\Ebarp} \regimei.
\end{align}
\item $\cdbar<\cd< \cdtil(\ca)$: The defense cost is relatively high in $\gamen$, but relatively low in $\games$. We denote the set of $\(\ca, \cd\)$ that satisfy this condition as $\regimem$ (\emph{medium} cost). That is,
\begin{align}\label{M_set}
\regimem \deleq \left\{\(\ca, \cd\)|\cdbar<\cd< \cdtil(\ca) \right\}=\cup_{i=1}^{\Ebarp} \(\regimesi \setminus \regimei\).
\end{align}
\item $\cd>\cdtil(\ca)$: The defense cost is relatively high in both $\gamen$ and $\games$. We denote the set of $\(\ca, \cd\)$ that satisfy this condition as $\regimeh$ (\emph{high} cost). That is,
\begin{align*}
\regimeh \deleq \left\{\(\ca, \cd\)|\cd>\cdtil(\ca)\right\}=\cup_{j=1}^{\Ebarp} \regimesj.
\end{align*}
\end{itemize}

We next compare the properties of NE and SPE for cost parameters in each set based on Theorems \ref{attacker_strategy} and \ref{theorem:SPE}, and Propositions \ref{utility}.
\begin{itemize}
\item Set $\regimel$:

\noindent\emph{Attacker}: In $\gamen$, the total attack probability is nonzero but smaller than 1, whereas in $\games$, the attacker is fully deterred. The attacker's equilibrium utility is identical in both games, i.e., $\Ua=\Uas$.

\noindent\emph{Defender}: The defender chooses identical equilibrium security effort in both games, i.e. $\pwe=\pswe$, but obtains a higher utility in $\games$ in comparison to that in $\gamen$, i.e., $\Ud<\Uds$.
\item Set $\regimem$:

\noindent\emph{Attacker}: In $\gamen$, the attacker conducts an attack with probability 1, whereas in $\games$ the attacker is fully deterred. The attacker's equilibrium utility is lower in $\games$ in comparison to that in $\gamen$, i.e., $\Ua>\Uas$. 

\noindent\emph{Defender}: The defender secures each vulnerable facility with a strictly higher level of effort in $\games$ than in $\gamen$, i.e. $\psewe>\pewe$ for each vulnerable facility $\e \in \{\E |\Ce-\ca>\Czero\}$. The defender's equilibrium utility is higher in $\games$ in comparison to that in $\gamen$, i.e., $\Ud<\Uds$. 
\item Set $\regimeh$:

\noindent\emph{Attacker}: In both games, the attacker conducts an attack with probability 1, and obtains identical utilities, i.e. $\Ua=\Uas$.

\noindent\emph{Defender}: The defender chooses identical equilibrium security effort in both games, i.e., $\pwe=\pswe$, and obtains identical utilities, i.e. $\Ud=\Uds$.
\end{itemize}

Importantly, the key difference between NE and SPE comes from the fact that in $\games$, the defender as the leading player is able to influence the attacker's strategy in her favor. 
Hence, when the defense cost is relatively medium or low (both sets $\regimem$ and $\regimel$), the defender can proactively secure all vulnerable facilities with the threshold effort to fully deter the attack, which results in a higher defender utility in $\games$ than in $\gamen$. Thus, we say the defender has the first-mover advantage when the cost parameters lie in the set $\regimem$ or $\regimel$. However, the reason behind the first-mover advantage differs in each set:
\begin{itemize}
\item In set $\regimem$, the defender needs to proactively secure all vulnerable facilities with strictly higher effort in $\games$ than that in $\gamen$ to fully deter the attacker. 
\item In set $\regimel$, the defender secures facilities in $\games$ with the same level of effort as that in $\gamen$, and the attacker is still deterred with probability 1. 
\end{itemize}
On the other hand, in set $\regimeh$, the defense cost is so high that the defender is not able to secure all targeted facilities with an adequately high level of security effort. Thus, the attacker conducts an attack with probability 1 in both games, and the defender no longer has first-mover advantage.

\begin{figure}[htp]
\centering
\includegraphics[width=0.5\textwidth]{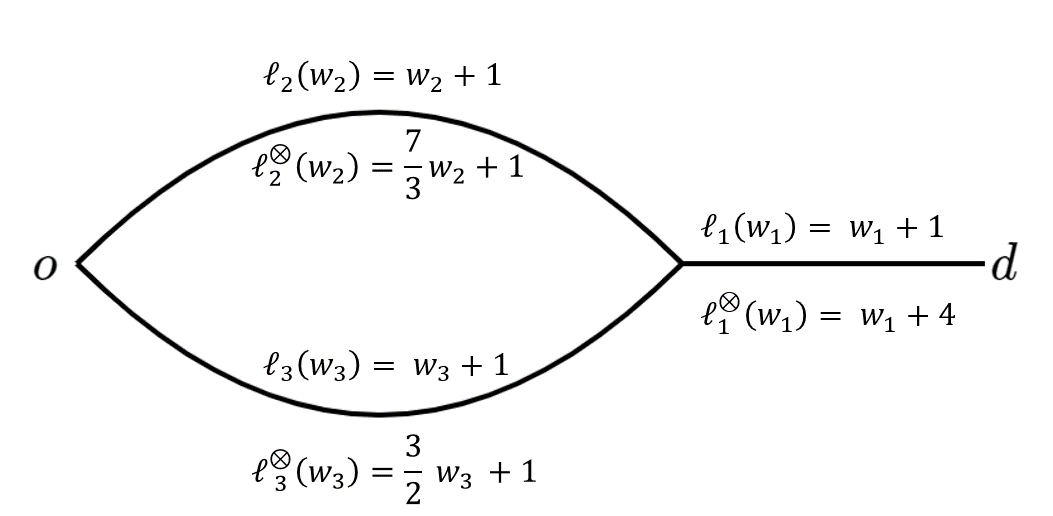}
\caption{Three edge network}
\label{three_facility}
\end{figure}

\begin{figure}[H]
\centering
\begin{subfigure}{0.32 \textwidth}
\includegraphics[width=\textwidth]{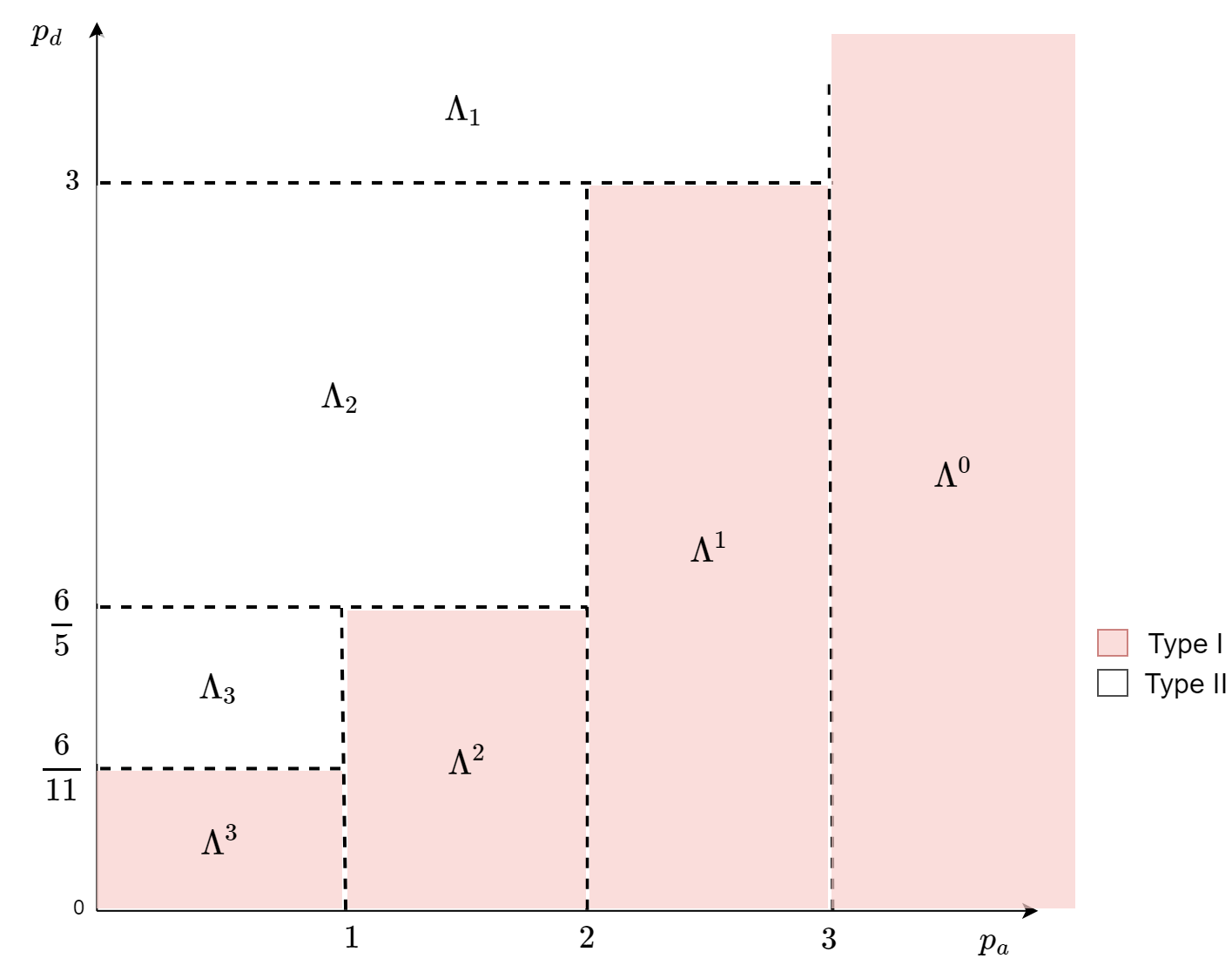}
\caption{}
\end{subfigure}
\begin{subfigure}{0.32 \textwidth}
\includegraphics[width=\textwidth]{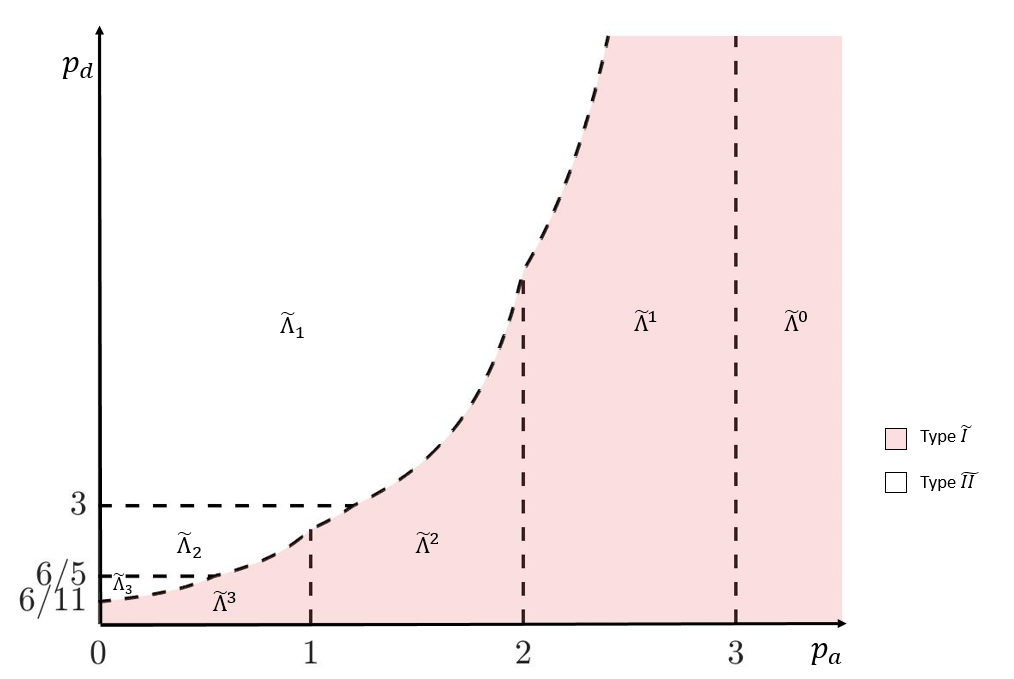}
\caption{}
\end{subfigure}
\begin{subfigure}{0.32 \textwidth}
\includegraphics[width=\textwidth]{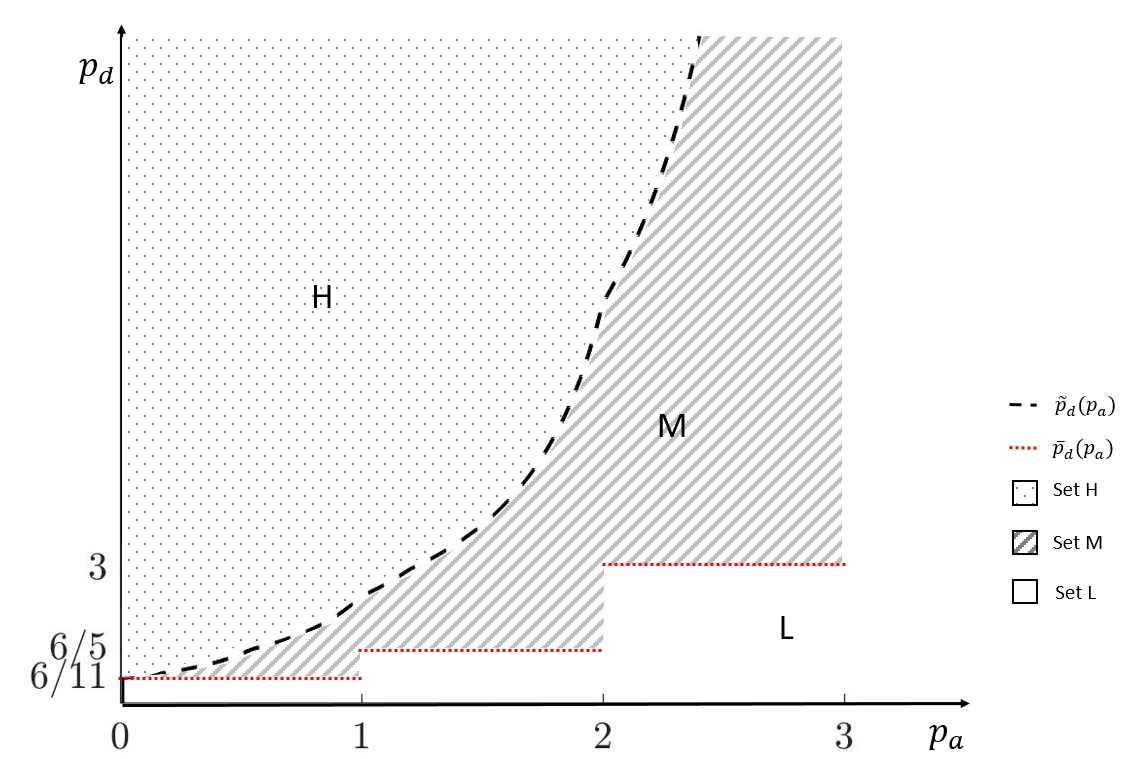}
\caption{}
\end{subfigure}
\caption{(a) Regimes of NE in $\gamen$, (b) Regimes of SPE in $\games$, (c) Comparison of NE and SPE. }
\label{ten_regime}
\end{figure}

Finally, for the sake of illustration, we compute the parameter sets $L$, $M$, and $H$ for transportation network with three facilities (edges); see Fig. \ref{three_facility}. If an edge $\e \in \E$ is not damaged, then the cost function is $\ell_e(w_e)$, which increases in the edge load $w_e$. If edge $\e$ is successfully compromised by the attacker, then the cost function changes to $\ell_\e^{\otimes}(w_e)$, which is higher than $\ell_e(w_e)$ for any edge load $w_e>0$. 
The network faces a set of non-atomic travelers with total demand $D=10$. We define the usage cost in this case as the average cost of travelers in Wardrop equilibrium \cite{correa2011wardrop}. Therefore, the usage costs corresponding to attacks to different edges are $C_1=20$, $C_2=19$, $C_3=18$ and the pre-attack usage cost is $\Czero=17$. From \eqref{order}, $\Ebarp=3$, and $\Ebar_{(1)}=\{e_1\}$, $\Ebar_{(2)}=\{e_2\}$ and $\Ebar_{(3)}=\{e_3\}$. 
In Fig. \ref{ten_regime}, we illustrate the regimes of both $\gamen$ and $\games$, and the three sets $H$, $M$, and $L$ distinguished by the thresholds $\cdbar$ and $\cdtil(\ca)$. 

\section{Model Extensions and Dynamic Aspects}\label{example_sec}
In this section, we first discuss how relaxing our modeling assumptions influence our main results. Next we introduce a dynamic setup in which the users of the infrastructure system face uncertainty about the outcome of attacker-defender interaction (i.e., identity of the compromised facility), and follow a repeated learning procedure to make their usage decisions.

\subsection{Relaxing Model Assumptions}\label{extension}
Our discussion centers around extending our results when the following modeling aspects are included:  facility-dependent cost parameters, less than perfect defense, and attacker's ability to target multiple facilities. 

\begin{enumerate}
\item \emph{Facility-dependent attack and defense costs.} 

Our techniques for equilibrium characterization of games $\gamen$ and $\games$ --- as presented in Sections \ref{sec:generic_case} and \ref{sequential_section} respectively --- can be generalized to the case when attack/defense costs are non-homogeneous across facilities. We denote the attack (resp. defense) cost for facility $\e\in\E$ as $\cae$ (resp. $\cde$). However, an explicit characterization of equilibrium regimes in each game can be quite complicated due to the multidimensional nature of cost parameters. 

In normal form game $\gamen$, it is easy to show that the attacker's best response correspondence in Lemma \ref{only_attacked} holds except that the threshold attack probability for any facility $\e \in \Ebar$ now becomes $\cde/(\Ce-\Czero)$. The set of vulnerable facilities is given by $\{\E|\Ce-\cae>\Czero\}$. The attacker's equilibrium strategy is to order the facilities in decreasing order of $\Ce-\cae$, and target the facilities in this order each with the threshold probability until either all vulnerable facilities are targeted or the total probability of attack reaches 1. As in Theorem \ref{attacker_strategy}, the former case happens when the cost parameters lie in a type I regime, and the latter case happens for type II regimes, although the regime boundaries are more complicated to describe. In equilibrium, the defender chooses the security effort vector to ensure that the attacker is indifferent among choosing any of the pure actions that are in the support of equilibrium attack strategy. 

In the sequential game $\games$, Lemmas \ref{best_response_sequential} and \ref{zero_or_one} can be extended in a straightforward manner except that the threshold security effort for any vulnerable facility $\e \in \{\E|\Ce-\Czero>\cae\}$ is given by $\pebar=(\Ce-\cae-\Czero)/(\Ce-\Czero)$. The SPE for this general case can be obtained analogously to Theorem \ref{theorem:SPE}, i.e. comparing the defender's utility of either securing all vulnerable facilities with the threshold effort to fully deter the attack, or choosing a strategy that is identical to that in $\gamen$. These cases happen when the cost parameters lie in (suitably defined) Type $\typeone$ and Type $\typetwo$ regimes, respectively. The main conclusion of our analysis also holds: the defender obtains a higher utility by proactively defending all vulnerable facilities when the facility-dependent cost parameters lie in type $\typeone$ regimes. 

\item \emph{Less than perfect defense in addition to facility-dependent cost parameters.}

Now consider that the defense on each facility is only successful with probability $\gamma \in (0,1)$, which is an exogenous technological parameter. For any security effort vector $\rho$, the actual probability that a facility $\e$ is not compromised when targeted by the attacker is $\gamma \rho_\e$. Again our results on NE and SPE in Sec. \ref{sec:generic_case} -- Sec. \ref{sequential_section} can be readily extended to this case. However, the expressions for thresholds for attack probability and security effort level need to be modified. 
In particular, for $\gamen$, in Lemma \ref{only_attacked}, the threshold attack probability on any facility $\e \in \Ebar$ is $\cde/\gamma(\Ce-\Czero)$. For $\games$, the threshold security effort $\pebar$ for any vulnerable facility $\e \in \{\E|\Ce-\Czero>\cde\}$ is $(\Ce-\cae-\Czero)/\gamma(\Ce-\Czero)$. If this threshold is higher than 1 for a particular facility, then the defender is not able to deter the attack from targeting it.

\item \emph{Attacker's ability to target multiple facilities.} 

If the attacker is not constrained to targeting a single facility, his pure strategy set would be $\Sa=2^{\E}$. Then for a pure strategy profile $\(\sd, \sa\)$, the set of compromised facilities is given by $\sa \setminus \sd$, and the usage cost $C_{\sa \setminus \sd}$.
Unfortunately, our approach cannot be straightforwardly applied to this case. This is because the mixed strategies cannot be equivalently represented as probability vectors with elements representing the probability of each facility being targeted or secured. In fact, for a given attacker's strategy, one can find two feasible defender's mixed strategies that induce an identical security effort vector, but result in different players utilities. Hence, the problem of characterizing defender's equilibrium strategies cannot be reduced to characterizing the equilibrium security effort on each facility. 
Instead, one would need to account for the attack/defense probabilities on all the subsets of facilities in $\E$. This problem is beyond the scope of our paper, although a related work \cite{dahan2015network} has made some progress in this regard.
\end{enumerate}

Finally, we briefly comment on the model where all the three aspects are included. So long as players' strategy sets are comprised of mixed strategies, the defender's equilibrium utility in $\games$ must be higher or equal to that in $\gamen$. This is because in $\games$, the defender can always choose the same strategy as that in NE to achieve a utility that is no less than that in $\gamen$. Moreover, one can show the existence of cost parameters such that the defender has strictly higher equilibrium utility in SPE than in NE. In particular, consider that the attacker's cost parameters $\(\cae\)_{\e\in\E}$ in this game are such that there is only one vulnerable facility $\ebar \in \E$ such that $C_{\ebar}-\Czero>p_{a, \ebar}$, and the threshold effort on that facility $\widehat{\rho}_{\ebar}=\(C_{\ebar}-p_{a,\ebar}-\Czero\)/\gamma(C_{\ebar}-\Czero)<1$. In this case, if the defense cost $p_{d, \ebar}$ is sufficiently low, then by proactively securing the facility $\ebar$ with the threshold effort $\widehat{\rho}_{\ebar}$, the defender can deter the attack completely and obtain a strictly higher utility in $\games$ than that in $\gamen$. Thus, for such cost parameters, the defender gets the first mover advantage in equilibrium.

\subsection{Rational Learning Dynamics}\label{example_dynamic}
We now discuss an approach for analyzing the dynamics of usage cost after a security attack. Recall that the attacker-defender model enables us to evaluate the vulnerability of individual facilities to a strategic attack for the purpose of prioritizing defense investments. One can view this model as a way to determine the set of possible post-attack states, denoted $\s \in \S \deleq \E\cup \{\emptyset\}$. In particular, we consider situations in which the distribution of the system state, denoted $\ps \in \Delta(\S)$, is determined by an equilibrium of attacker-defender game ($\gamen$ or $\games$). In $\gamen$, for each $\s \in \S$, the probability $\ps(\s)$ is given as follows:
\begin{align}\label{ps}
\ps(\s)=\left\{
\begin{array}{ll}
\siganwe(\e) \cdot (1-\pewe), \quad & \quad \text{if $\s=\e$,}\\
1-\sum_{\e \in \E} \ps(\e), \quad & \quad \text{if $\s=\emptyset$.}
\end{array}
\right.
\end{align}
For $\games$ the probability distribution $\ps$ can be analogously defined in terms of $\sigaswe$ and $\pswe$. 

Let the realized state be $\s=\e$, i.e., the facility $\e \in\E$ is compromised by the attacker. If this information is known perfectly to all the users immediately after the attack, they can shift their usage choices in accordance to the new state. Then the cost resulting from the users' choices indeed corresponds to the usage cost $C_s=C_{e}$, which governs the realized payoffs of both attacker and defender. However, from our results (Theorems \ref{attacker_strategy} and \ref{theorem:SPE}), it is apparent that the support of equilibrium player strategies (and hence the support of $\ps$) can be quite large. Due to inherent limitations in perfectly diagnosing the location of attack, in some situations, the users may not have full knowledge of the realized state. Then, the issues of how users with imperfect information make their decisions in a repeated learning setup, and whether or not the long-run usage cost converges to the actual cost $C_{e}$ become relevant. 

To contextualize the above issues, consider the situation in which a transportation system is targeted by an external hacker, and that the operation of a single facility is compromised. Furthermore, the nature of attack is such that travelers are not able to immediately know the identity of this facility. This situation can arise when the diagnosis of attack and/or dissemination of information about the attack is imperfect. Examples include cyber-security attacks to transportation facilities that can result in hard-to-detect effects such as compromised traffic signals of a major intersection, or tampering of controllers governing the access to a busy freeway corridor. Then, one can study the problem of learning by rational but imperfectly informed travelers using a repeated routing game model. We now discuss the basic ideas behind the study of this problem. A more rigorous treatment is part of our ongoing work, and will be detailed in a subsequent paper.

Let the stages of our repeated routing game be denoted as $\t \in \T =\{1, 2, \dots\}$. In this game, travelers are imperfectly informed about the network state. In particular, in each stage $\t\in \T$, they maintain a belief about the state $\thetat$. The initial belief $\thetazero$ can be different from the prior state distribution $\ps$. However, we require that $\thetazero$ is absolutely continuous with respect to $\ps$ (\cite{kalai1993subjective}):
\[\forall \s \in \S, \quad \theta(\s)>0, \quad \Rightarrow \quad \thetazero(\s)>0. \]
That is, the initial belief of travelers does not rule out any possible state. 

The solution concept we use for this repeated game is \textsl{Markov-perfect Equilibrium} (see \cite{maskin2001markov}), in which travelers use routes with the smallest expected cost based on the belief in each stage. Equivalently, the equilibrium routing strategy in stage $\t$ is a Wardrop equilibrium of the stage game with belief $\thetat$ (\cite{correa2011wardrop}). We also consider that at the end of each stage, travelers receive noisy information of the realized costs on routes that are taken. However, no information is available for routes that are not chosen by any traveler. Based on the received information, travelers update their belief of the state using Bayes' rule. 

We note that numerous learning schemes have been studied in the literature; for e.g. fictitious play (\cite{brown1951iterative}, \cite{fudenberg1995consistency}, and \cite{hofbauer2002global});  reinforcement learning (\cite{beggs2005convergence}, \cite{cominetti2012adaptive} and \cite{cominetti2010payoff}), and regret minimizations (\cite{blum2006routing} and \cite{marden2007regret}). These learning schemes typically assume that strategies in each stage are determined by a certain function of the history payoff or actions. To explain the learning dynamics
in our set-up we consider that in each stage travelers are rational, and they aim to maximize the payoff myopically based on their current belief about other travelers' strategies. The players update their beliefs based on observed actions on the play-path. This so-called rational learning dynamics has been investigated by \cite{battigalli1992learning}, \cite{fudenberg1995learning}, \cite{kalai1993subjective}, and \cite{kalai2015learning}. Our model is different from the ones in literature in that travelers are uncertain about the payoff functions, but correctly anticipate the opponents' strategies. Additionally, the information of the payoff in each stage is noisy and limited (only the realized costs on the taken routes are known). 

The game can be understood easily via an example of a transportation network in Fig. \ref{three_facility}. In each stage $\t$, travelers with inelastic demand $\D$ choose route $r_1$ ($e_2-e_1$) or route $r_2$ ($e_3-e_1$). We denote the equilibrium routing strategy in stage $\t$ as $q^{\t*}(\thetat)=\(q^{\t*}_r(\thetat)\)_{r \in \{r_1, r_2\}}$, where $q^{\t*}_r(\thetat)$ is the demand of travelers using route $r$ given the belief $\thetat$. Hence, aggregate flow on edge $e_2$ (resp. $e_3$) is $w_2^{t*}(\thetat)=q_1^{t*}(\thetat)$ (resp. $w_3^{t*}(\thetat)=q_2^{t*}(\thetat)$), and the aggregate flow on edge $e_1$ is $w_1^{t*}(\thetat)=\D$. Each stage game is a congestion game, and hence admits a potential function. The equilibrium routing strategy $q^{t*}(\thetat)$ can be computed efficiently for this game. Moreover, in each stage, the equilibrium is essentially unique in that the equilibrium edge load is unique for a given belief (\cite{sandholm2001potential}). 

The realized cost on each edge $\e \in \E$, denoted $c_e^s(w_e^{t*}(\thetat))$, equals to the cost (shown in Fig. \ref{three_facility} for the example network) plus a random variable $\epe$:
\begin{align}\label{epe}
c_e^s(q^{t*}(\thetat))=\left\{
\begin{array}{ll}
\ell_\e^\otimes(w_e^{t*}(\thetat))+\epe, & \quad \text{if $\s=\e$,}\\
\ell_\e(w_e^{t*}(\thetat))+\epe, & \quad \text{otherwise.}
\end{array}
\right.
\end{align}

We illustrate two cases that can arise in rational learning:

\begin{itemize}
\item \emph{Long-run usage cost equals to $\Cs$ for any $\s \in \{e_1, e_2, e_3, \emptyset\}$.}

Consider the case where the initial belief is $\ps(e_1)=1/12$, $\ps(e_2)=1/3$, $\ps(e_3)=1/12$, $\ps(\emptyset)=1/2$ (The initial belief can be any probability vector which satisfies the continuity assumption). For any $\e \in \E$, the random variable $\epe$ in \eqref{epe} is distributed as $U[-3, 3]$. The total demand $\D=10$. Fig. \ref{theta_1}--\ref{theta_4} show how the belief of each state evolves. We see that eventually travelers learn the true state, and hence the long-run usage cost converges to the actual post-attack usage cost $C_\s$, even though initially all travelers are imperfectly informed about the state. 
\begin{figure}[htp]
\centering
    \begin{subfigure}[b]{0.40 \textwidth}
        \includegraphics[width=\textwidth]{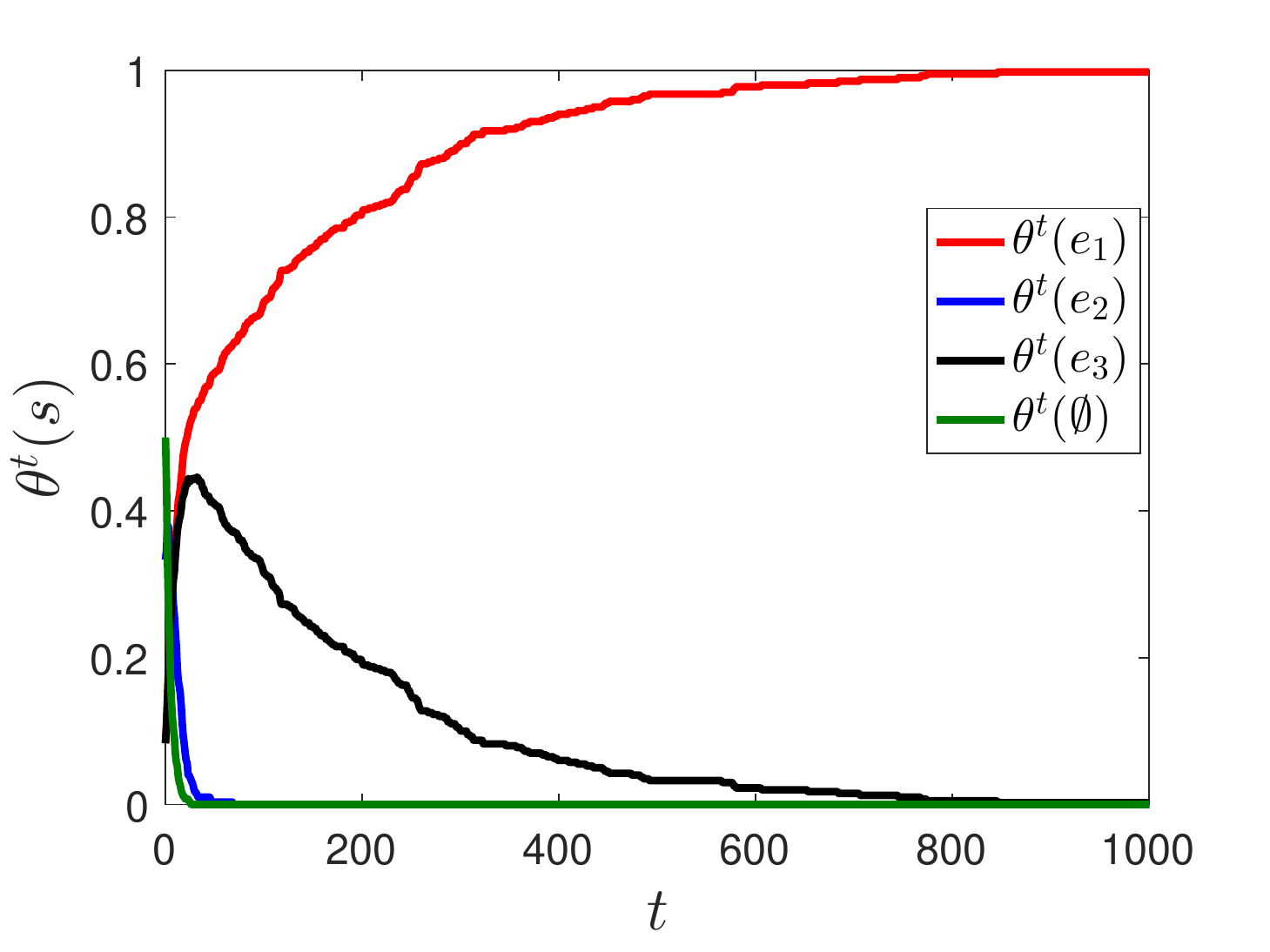}
        \caption{$\s=e_1$.}
        \label{theta_1}
    \end{subfigure}
~
\centering
	\begin{subfigure}[b]{0.40\textwidth}
        \includegraphics[width=\textwidth]{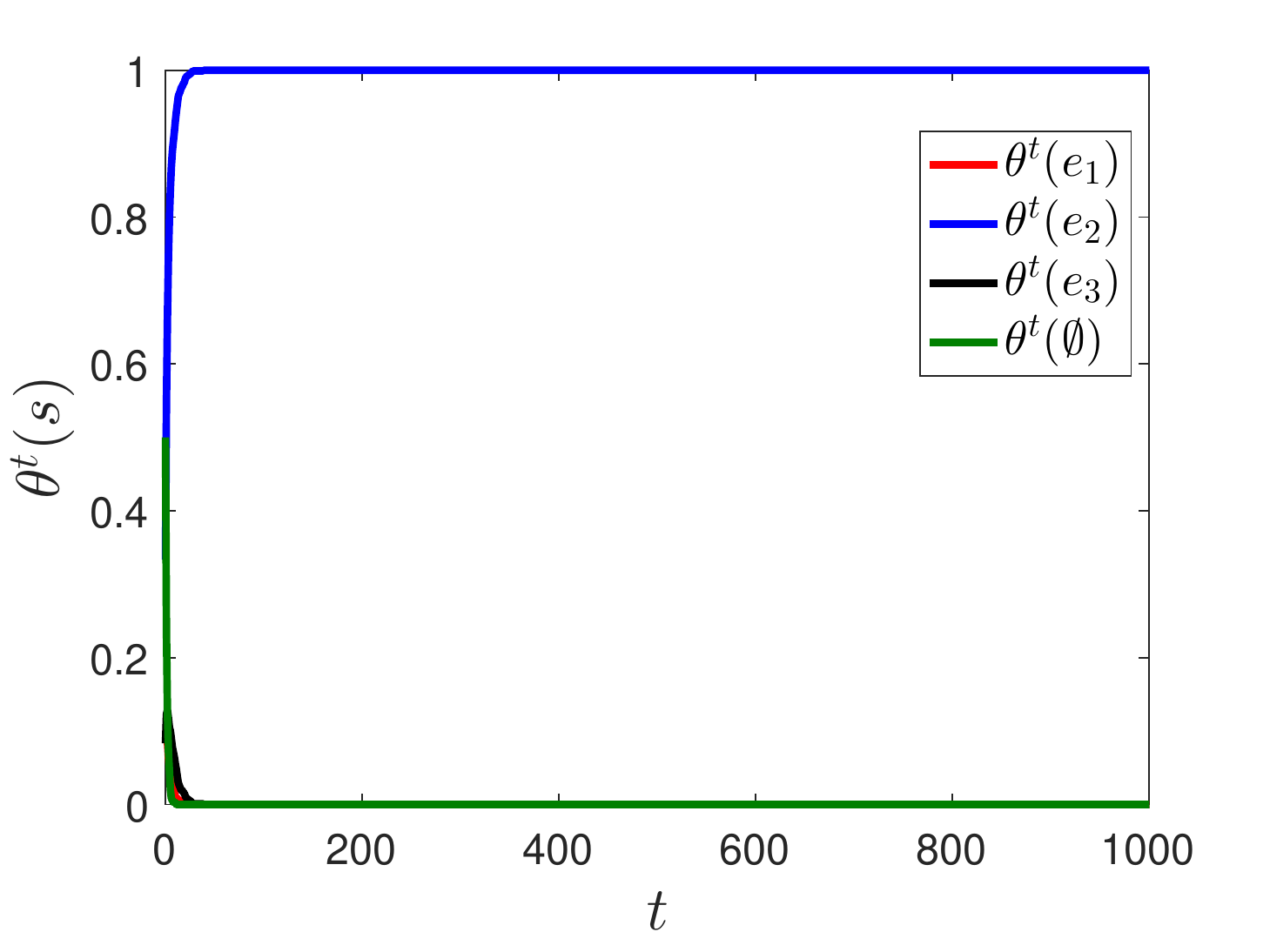}
        \caption{$\s=e_2$.}
        \label{theta_2}
    \end{subfigure}\\
   \begin{subfigure}[b]{0.40\textwidth}
        \includegraphics[width=\textwidth]{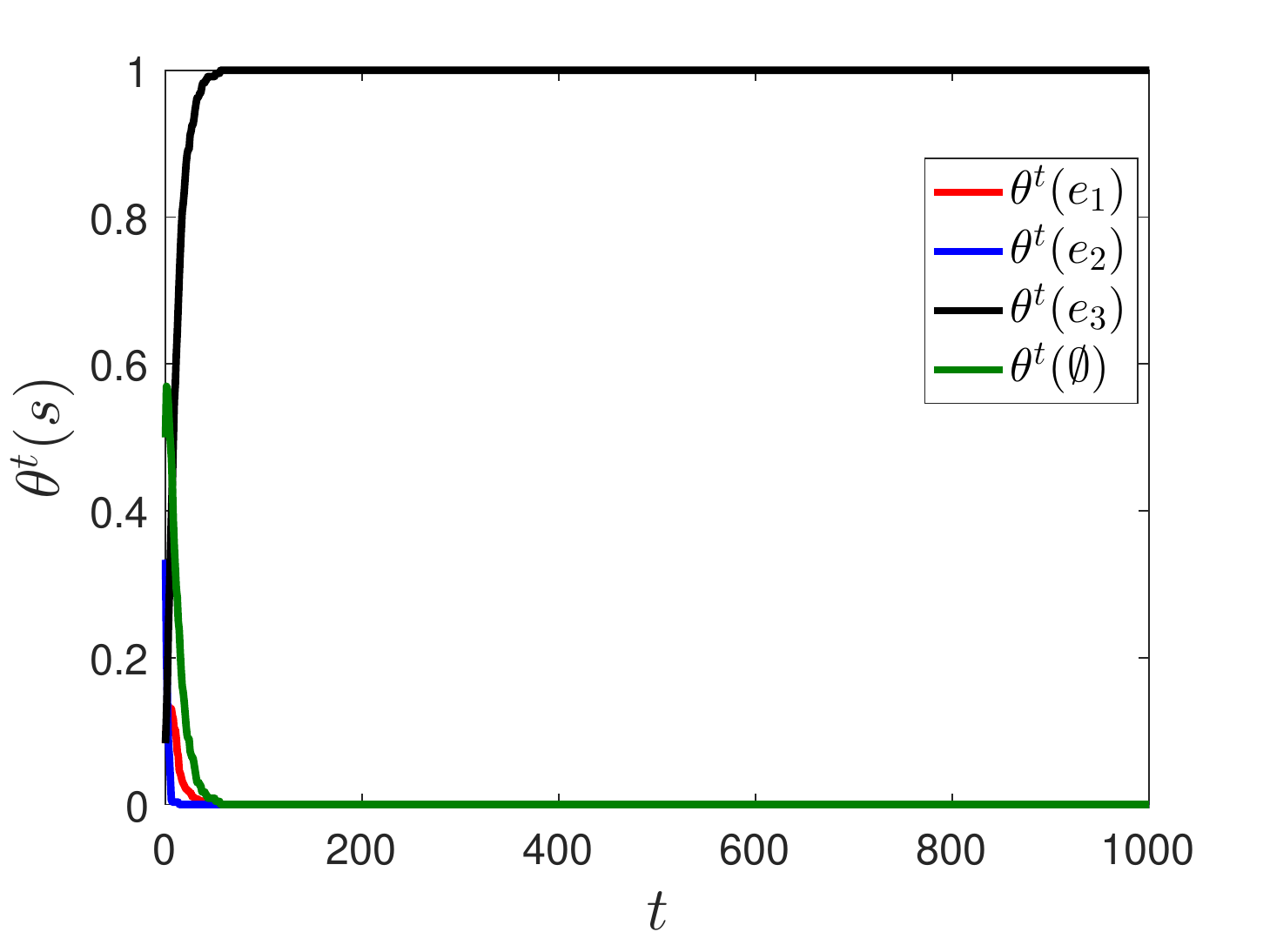}
        \caption{$\s=e_3$.}
        \label{theta_3}
    \end{subfigure}
~
\centering
	\begin{subfigure}[b]{0.40\textwidth}
        \includegraphics[width=\textwidth]{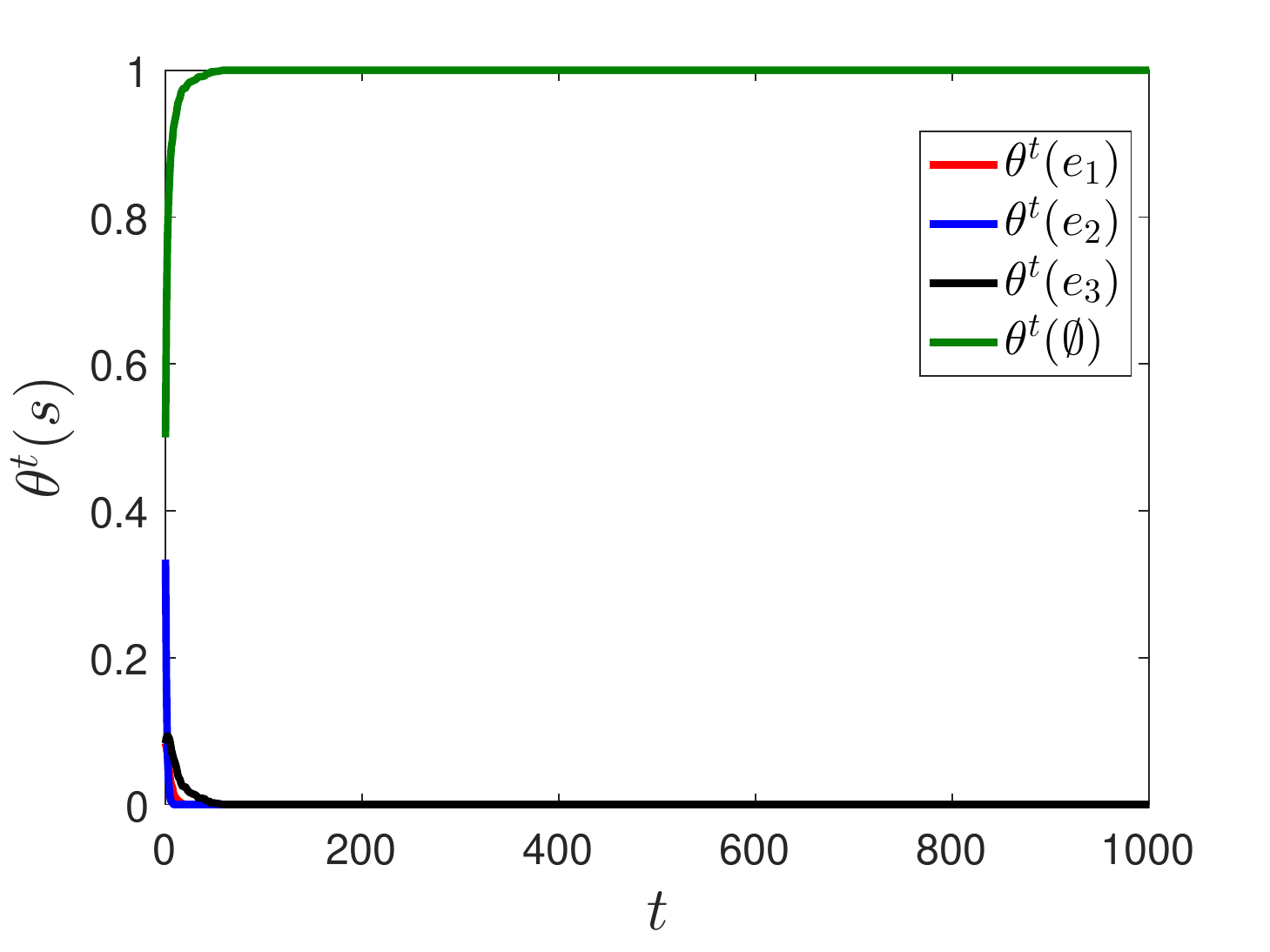}
        \caption{$\s=\emptyset$.}
        \label{theta_4}
    \end{subfigure}
    \caption{Rational learning leads to the usage cost of the true state.}
\label{fig:correct_prior_converge}
\end{figure}

\item \emph{Long-run usage cost is higher than $\Cs$. }

Consider the case when, as a result of attack on edge $e_2$, the cost function on $\e_2$ changes to $\ell_2^\otimes(w_2)=7/3w_2+50$. The total demand is $\D=5$, and the initial belief is $\ps(e_1)=1/12$, $\ps(e_2)=1/3$, $\ps(e_3)=1/12$, $\ps(\emptyset)=1/2$. Starting from this initial belief, travelers exclusively take route $r_2$, and hence they do not obtain any information about $e_2$. Even when the realized state is $\s=\emptyset$, travelers end up repeatedly taking $r_2$ as if $e_2$ is compromised. Thus, the long-run average cost is $C_{e_2}$, which is higher than the cost corresponding to the true state $\Czero$. Therefore, rational learning dynamics can lead to long-run inefficiency. We illustrate the equilibrium routing strategies and beliefs in each stage in Fig. \ref{q_self_confirm} and Fig. \ref{theta_self_confirm} respectively. 
\begin{figure}[htp]
    \begin{subfigure}[b]{0.40 \textwidth}
        \includegraphics[width=\textwidth]{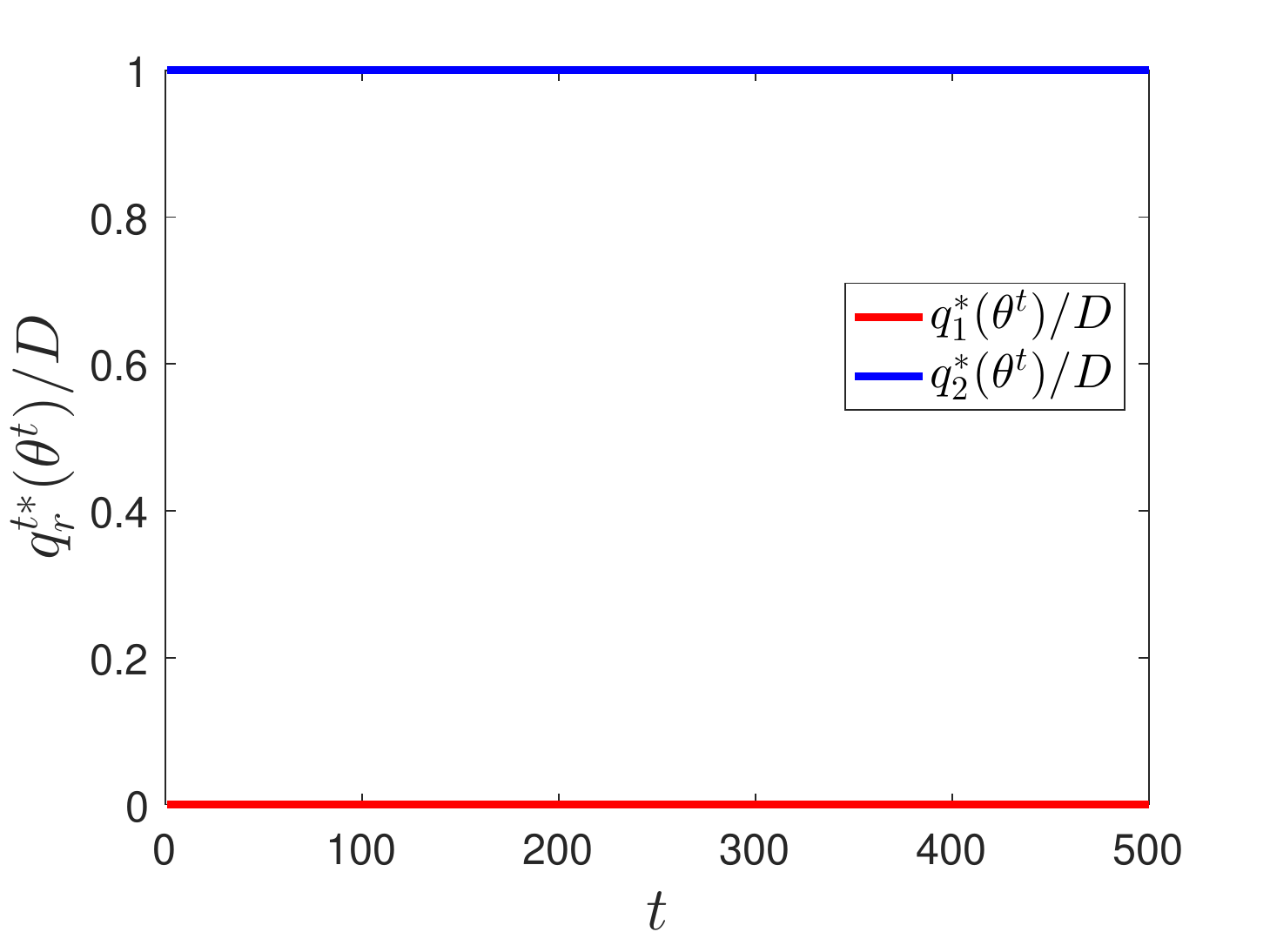}
        \caption{}
        \label{q_self_confirm}
    \end{subfigure}
~
\centering
	\begin{subfigure}[b]{0.40\textwidth}
        \includegraphics[width=\textwidth]{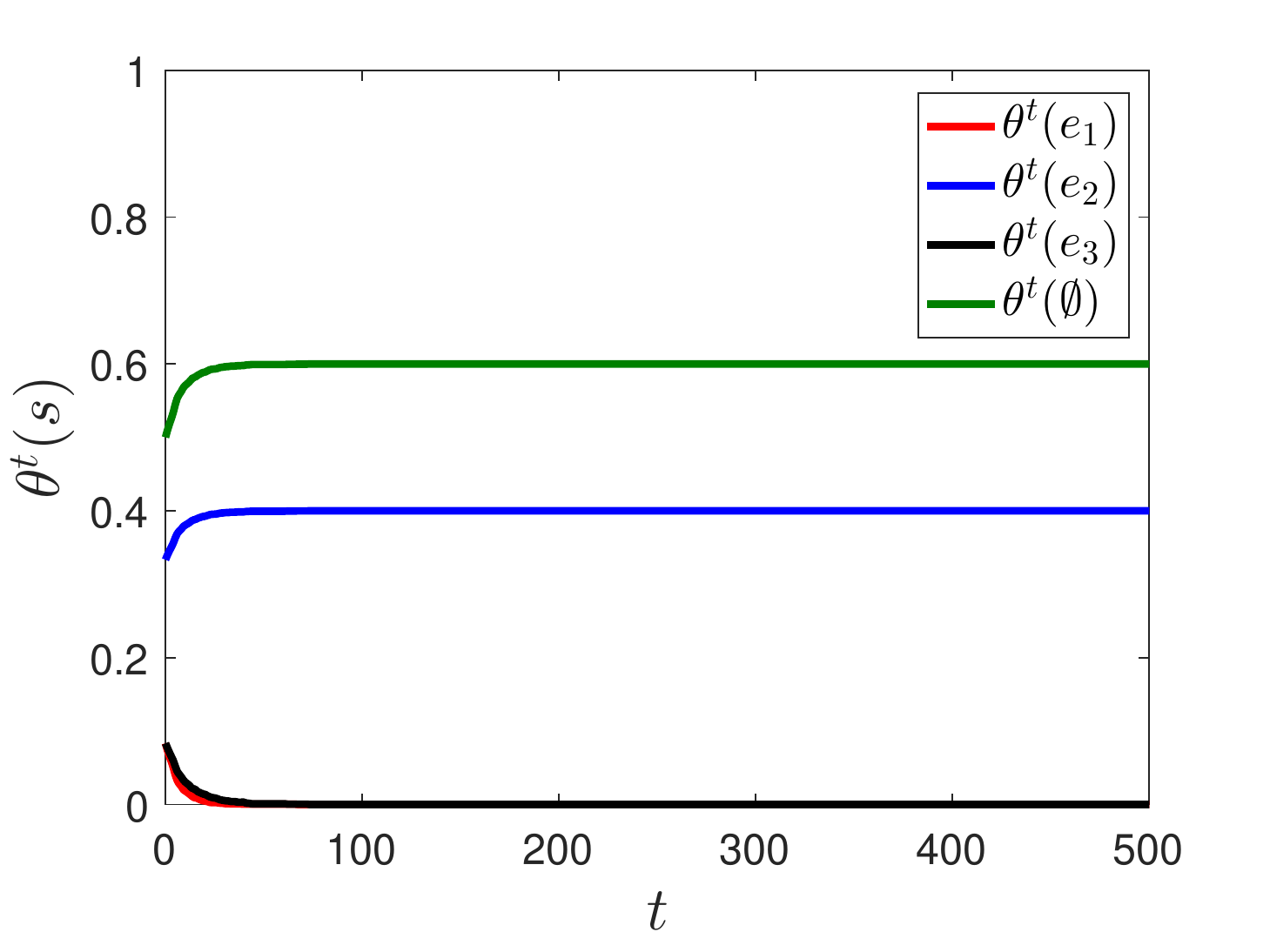}
        \caption{}
        \label{theta_self_confirm}
    \end{subfigure}
    \caption{Learning leads to long-run inefficiency $\sran=\emptyset$: (a) Equilibrium routing strategies; (b) Beliefs. }
\label{fig:self_confirm}
\end{figure}
\end{itemize}
These cases illustrate that if the post-attack state is not perfectly known by the users of the system, then the cost experienced by the users depend on the learning dynamics induced by the repeated play of rational users. Particularly, the learning dynamics can induce a higher usage cost in the long-run in comparison to the cost corresponding to the true state. Following previously known results \cite{fudenberg1993steady}, one can argue that if sufficient amount of ``off-equilibrium'' experiments are conducted by travelers, then the learning will converge to Wardrop equilibrium with the true state. However, such experiments are in general not costless. 

As a final remark, we note another implication of proactive defense strategy in ranges of attack/ defense cost parameters where the first-mover advantage holds. In particular, when the cost parameters are in the sets $L$ and $M$ as given in \eqref{L_set}-\eqref{M_set}, the attack is completely deterred in the sequential game $\games$ and there is no uncertainty in the realized state. In such a situation,
one does not need to consider uncertainty in the travelers' belief about the true state and issue of long-run inefficiency due to learning behavior does not arise.

\section*{Acknowledgments}
We are sincerely thankful to Prof. Georges Zaccour and two anonymous referees whose constructive comments helped us to improve our initial manuscript. We thank seminar participants at MIT, HEC Montreal, University of Pennsylvania, and NYU Abu Dhabi for helpful comments. The authors are grateful to Professors Alexandre Bayen, Patrick Jaillet, Karl Johansson, Patrick Loiseau, Samer Madanat, Hani Mahmassani, Asu Ozdaglar, Galina Schwartz, Demos Teneketzis, Rakesh Vohra, Dan Work, Georges Zaccour for insightful comments and discussions in the early phase of this research. This work was supported in part by Singapore-MIT Alliance for Research and Technology (SMART) Center for Future Mobility (FM), NSF grant CNS 1239054, NSF CAREER award CNS 1453126.

\bibliographystyle{plain}

\bibliography{library}

\newpage
\begin{appendix}
\section{Proofs of Section \ref{Sec:attack-defend}}\label{appendix_three}
\noindent\emph{Proof of Lemma \ref{lemma:stra_construct}.}
We first show that the strategy in \eqref{stra_construct} is feasible. Since $\porderone \leq 1$, and for any $i=1, \dots, m-1$, $\porderi-\porderipone>0$, $\sigdn(\sd)$ is non-negative for any $\sd \in \Sd$. Additionally, 
\begin{align*}
\sum_{\sd \in \Sd}\sigdn(\sd)&=\sigdn\(\emptyset\)+\sum_{i=1}^{m-1} \sigdn\(\left\{\e \in \E| \pee \geq \porderi\right\}\)+\sigdn\(\left\{\e \in \E| \pee \geq \porderm \right\}\)\\
&=\(1-\porderone\)+\sum_{i=1}^{m-1}\(\porderi-\porderipone\)+\porderm\\
&=1-\porderone+\porderone-\porderm+\porderm\\
&=1.
\end{align*}
Thus, $\sigdn$ in \eqref{stra_construct} is a feasible strategy of the defender. Now we check that $\sigdn$ in \eqref{stra_construct} indeed induces $\rho$. Consider any $\e \in \E$ such that $\pee=0$. Then, since $\e \notin \left\{\E|\pee \geq \porderi \right\}$ for any $i=1, \dots, m$, and $e \notin \emptyset$, for any $\sd \ni \e$, we must have $\sigdn(\sd)=0$. Thus, $\sum_{\sd \ni \e} \sigdn(\sd)=0=\pee$. Finally, for any $j=1, \dots, m$, consider any $\e \in \E$, where $\pee=\rho_{(j)}$:
\begin{align*}
\sum_{\sd \ni \e} \sigdn(\sd)=\sum_{i=j}^{m} \sigdn\(\left\{\e \in \E| \pee \geq \porderi\right\}\)=\rho_{(j)}.
\end{align*} 
Therefore, $\sigdn$ in \eqref{stra_construct} induces $\rho$. 
 \qed
 \vspace{0.3cm}
 
%

\noindent\emph{Proof of Proposition \ref{strict_dominated}.}
We prove the result by the principal of iterated dominance. We first show that any $\sd$ such that $\sd \nsubseteq \Ebar$ is strictly dominated by the strategy $\sdp =\sd \cap \Ebar$. Consider any pure strategy of the attacker, $\sa \in \E$, the utilities of the defender with strategy $\sd$ and $\sdp$ are as follows:
\begin{align*}
\ud(\sd, \sa)&=-C(\sd, \sa)-|\sd| \cd=-C(\sd, \sa)-(|\sdp|+|\sd \setminus \Ebar|) \cd, \\
\ud(\sdp, \sa)&=-C(\sdp, \sa)-|\sdp| \cd.
\end{align*}
If $\sa \in \Ebar$ or $\sa \notin \sd$ or $\sa=\emptyset$, then $C(\sd, \sa)=C(\sdp, \sa)$, and thus $\Ud(\sd, \sa)<\Ud(\sdp, \sa)$. If $\sa=\e \in \sd \setminus \Ebar$, then $\e \notin \Ebar$, and $\Ce \leq \Czero$. We have $C(\sd, \sa)=\Czero \geq \Ce=C(\sdp, \sa)$, and thus $\Ud(\sdp, \sa) \geq -C(\sd, \sa)-|\sdp| \cd> \Ud(\sd, \sa)$. Therefore, any $\sd$ such that $\sd \nsubseteq \Ebar$ is a strictly dominated strategy. Hence, in $\gamen$, any equilibrium strategy of the defender satisfies $\sigdnwe(\sd)=0$. From \eqref{eq:ped}, we know that $\pewe=0$ for any $\e \in \E\setminus \Ebar$. 

We denote the set of defender's pure strategies that are not strictly dominated as $\Sdbar=\{\sd | \sd \subseteq \Ebar\}$. Consider any $\sd \in \Sdbar$, we show that any $\sa \in \E\setminus \Ebar$ is strictly dominated by strategy $\emptyset$. The utility functions of the attacker with strategy $\sa$ and $\emptyset$ are as follows:
\begin{align*}
\ua(\sd, \sa)&=C(\sd, \sa)-\ca,\\
\ua(\sd, \emptyset)&=C(\sd, \emptyset).
\end{align*}
Since $\sd \subseteq \Ebar$ and $\sa \in \E\setminus \Ebar$, $\sa \notin \sd$, thus $C(\sd, \sa)=C_{\sa}\leq \Czero$. However, $C(\sd, \emptyset) = \Czero$ and $\ca >0$. Therefore, $\Ua(\sd, \emptyset)> \Ua(\sd, \sa)$. Hence, any $\sa \in \E\setminus \Ebar$ is strictly dominated. Hence, in equilibrium, the probability of the attacker choosing facility $\e \in \E\setminus \Ebar$ is 0 in $\gamen$. 

We can analogously argue that in $\games$, $\psewe=0$ and $\sigaswe(\e, \psss)=0$ for any $\e \in \E\setminus \Ebar$. 
\qed
\section{Proofs of Section \ref{sec:generic_case}}\label{appendix_NE_proof}
\noindent\emph{Proof of Lemma \ref{zero_sum}.}
The utility functions of the attacker with strategy $\sigan$ in $\gamezero$ and $\gamen$ are related as follows:
\begin{align*}
\Uazero(\sigdn, \sigan) =\Ua(\sigdn, \sigan)+\mathbb{E}_{\sigdn}\left[|\sd|\right]\cdot \cd.
\end{align*}
Thus, for a given $\sigdn$, any $\sigan$ that maximizes $\Uazero(\sigdn, \sigan)$ also maximizes $\Ua(\sigdn, \sigan)$. So the set of best response strategies of the attacker in $\gamezero$ is identical to that in $\gamen$. Analogously, given any $\sigan$, the set of best response strategies of the defender in $\gamen$ is identical to that in $\gamezero$. Thus, $\gamezero$ and $\gamen$ are strategically equivalent, i.e. they have the same set of equilibrium strategy profiles. Using the interchangeability property of equilibria in zero-sum games, we directly 
obtain that for any $\sigdnwe \in \Sigma^{*}_d$ and any $\siganwe \in \Sigma^{*}_a$, $(\sigdnwe, \siganwe)$ is an equilibrium strategy profile.     \qed

\vspace{0.3cm}

\noindent\emph{Proof of Proposition \ref{opt_eq}.}
From Lemma \ref{zero_sum}, the set of attacker's equilibrium strategies $\Siganwe$ is the optimal solution of the following maximin problem: 
\begin{subequations}\label{maxmin}
\begin{align}
\max_{\sigan} \quad &\min_{\sd \in \Sd} \left\{\sum_{\e \in \Ebar} \(C(\sd, \e)+|\sd| \cd-\ca\)\cdot \sigan(\e)+ \(C(\sd, \emptyset) +|\sd| \cd\) \cdot \sigan(\emptyset) \right\}\notag \\
s.t. \quad &\sum_{\e \in \Ebar} \sigan(\e)+ \sigan(\emptyset)=1, \label{sum_sig}\\
&\sigan(\emptyset) \geq 0, \quad \sigan(\e) \geq 0, \quad \forall \e \in \Ebar. \label{eq:non-negative}
\end{align}
\end{subequations}
Given any $\sd \in \Sd$, we can express the objective fucntion in \eqref{maxmin} as follows:
\begin{align*}
&\sum_{\e \in \Ebar} \(C(\sd, \e)+|\sd| \cd-\ca\)\cdot \sigan(\e)+ \(C(\sd, \emptyset) +|\sd| \cd\) \cdot \sigan(\emptyset) \\
=&\sum_{\e \in \Ebar} \(C(\sd, \e)-\ca\)\cdot \sigan(\e)+ C(\sd, \emptyset) \sigan(\emptyset) + |\sd| \cd \cdot \(\sum_{\e \in \E}\sigan(\e)+\sigan(\emptyset)\)\\
\stackrel{\eqref{sum_sig}}{=}&\sum_{\e \in \Ebar}\sigan(\e)\cdot \( C(\sd, \e)-\ca\)+  |\sd| \cd+ \sigan(\emptyset) \cdot \Czero\\
=&\sum_{\e \in \Ebar}\sigan(\e) \cdot \(C(\sd, \e)-\ca\)+\cd \cdot \(\sum_{\e \in \Ebar} \mathbbm{1}\{\sd \ni \e\}\)+ \sigan(\emptyset) \cdot \Czero \\
=&\sum_{\e \in \Ebar}\(\sigan(\e) \cdot \(C(\sd, \e)-\ca\)+\cd \cdot \mathbbm{1}\{\sd \ni \e\}\)+\sigan(\emptyset) \cdot \Czero \\
\stackrel{\eqref{Ceq}}{=}&\sum_{\e \in \sd} \(\sigan(\e)\cdot  \(\Czero-\ca\)+\cd\)+\sum_{\e \in \Ebar \setminus \sd} \sigan(\e) \cdot \(\Ce-\ca\)+ \sigan(\emptyset) \cdot \Czero.
\end{align*}
Therefore, we can write:
\begin{align*}
&\min_{\sd \in \Sd}\left\{\sum_{\e \in \Ebar} \(C(\sd, \e)+|\sd| \cd-\ca\)\cdot \sigan(\e)+ \(C(\sd, \emptyset) +|\sd| \cd\) \cdot \sigan(\emptyset) \right\} \\
=&\min_{\sd \in \Sd}\left\{ \sum_{\e \in \sd} \(\sigan(\e)\cdot  \(\Czero-\ca\)+\cd\)+\sum_{\e \in \Ebar \setminus \sd} \sigan(\e) \cdot \(\Ce-\ca\)+ \sigan(\emptyset) \cdot \Czero \right\}\\
=&\sum_{\e \in \Ebar} \min \left\{\sigan(\e)\cdot \( \Czero-\ca\)+\cd,~ \sigan(\e) \cdot \(\Ce-\ca\)\right\}+\sigan(\emptyset) \cdot \Czero \\
=&V(\sigan).
\end{align*}
Thus \eqref{maxmin} is equivalent to \eqref{maxmin_min}, and $\Siganwe$ is the optimal solution set of \eqref{maxmin_min}

By introducing an $|\Ebar|$-dimensional variable $v=\(\ve\)_{\e \in \Ebar}$, \eqref{maxmin_min} can be changed to a linear optimization program \eqref{linear_maxmin}, and $\Siganwe$ is the optimal solution set of \eqref{linear_maxmin}. \qed

\vspace{0.3cm}
\noindent\emph{Proof of Lemma \ref{only_attacked}.}
We first argue that the defender's best response is in \eqref{best_response_normal}. For edge $\e \in \E$ such that $\sigan(\e)<\frac{\cd}{\Ce-\Czero}$, we have $\(\Czero-\Ce\)\sigan(\e)+\cd>0$. Since $\rho \in BR(\sigan)$ maximizes $\Ud(\sigdn, \sigan)$ as given in \eqref{Ud_rewrite}, $\rho_e$ must be 0. Additionally, Proposition \ref{strict_dominated} ensures that for any $\e \in \E \setminus \Ebar$, $\rho_e$ is 0. 

Analogously, if $\sigan(\e)>\frac{\cd}{\Ce-\Czero}$, then $\(\Czero-\Ce\)\sigan(\e)+\cd<0$, and the best response $\rho_e=1$. Finally, if $\sigan(\e)=\frac{\cd}{\Ce-\Czero}$, any $\rho_e \in [0, 1]$ can be a best response. 

We next prove \eqref{upper_bound}. 
We show that if a feasible $\sigan$ violates \eqref{sub:upper_bound}, i.e., there exists a facility, denoted $\ebarnew \in \Ebar$ such that $\sigan(\ebarnew) > \frac{\cd}{C_{\ebarnew}-\Czero}$, then $\sigan$ cannot be an equilibrium strategy. There are two cases:
\begin{enumerate}[label=(\alph*)]
\item There exists another facility $\ebarp \in \Ebar$ such that $\sigan(\ebarp) <\frac{\cd}{C_{\ebarp}-\Czero}$. Consider an attacker's strategy $\siganp$ defined as follows:
\begin{align*}
\siganp(\e)&=\sigan(\e), \quad \forall \e \in \Ebar \setminus \{\ebarnew, \ebarp\}, \quad \siganp(\emptyset)=\sigan(\emptyset),\\
\siganp(\ebarnew)&=\sigan(\ebarnew)-\epsilon, \\
\siganp(\ebarp)&=\sigan(\ebarp)+\epsilon, 
\end{align*}
where $\epsilon$ is a sufficiently small positive number so that $\siganp(\ebarnew) > \frac{\cd}{C_{\ebarnew}-\Czero}$ and $\siganp(\ebarp) <\frac{\cd}{C_{\ebarp}-\Czero}$. We obtain:
\begin{align*}
V(\siganp)-V(\sigan)\stackrel{}{=}\epsilon \(C_{\ebarp}-\Czero\)>0
\end{align*}
The last inequality holds from \eqref{Ebar} and $\ebarp \in \Ebar$. Therefore, $\sigan$ cannot be an attacker's equilibrium strategy.  
\item If there does not exist such $\ebarnew$ as defined in case (a), then for any $\e  \in \Ebar$, we have $\sigan(\e) \geq \frac{\cd}{C_{\e}-\Czero}$. Now consider $\siganp$ as follows:
\begin{align*}
\siganp(\e)&=\sigan(\e), \quad \forall \e \in \E \setminus \{\ebarnew\}, \\
\siganp(\ebarnew)&=\sigan(\ebarnew)-\epsilon, \\
\siganp(\emptyset)&=\sigan(\emptyset)+\epsilon,
\end{align*}
where $\epsilon$ is a sufficiently small positive number so that $\siganp(\ebarnew) > \frac{\cd}{C_{\ebarnew}-\Czero}$. We obtain:
\begin{align*}
V(\siganp)-V(\sigan)\stackrel{}{=}\epsilon \(\Czero-\(\Czero-\ca\)\)= \epsilon \ca>0.
\end{align*}
Therefore, $\sigan$ also cannot be an attacker's equilibrium strategy. 
\end{enumerate}
Thus, we can conclude from cases (a) and (b) that in equilibrium $\siganwe$ must satisfy \eqref{sub:upper_bound}. Additionally, from Proposition \ref{strict_dominated}, \eqref{zero_out} is also satisfied. \qed

\vspace{0.3cm}
\noindent\emph{Proof of Theorem \ref{attacker_strategy}.}
We first prove the attacker's equilibrium strategies in each regime. From Proposition \ref{opt_eq} and Lemma \ref{only_attacked}, we know that $\siganwe$ maximizes $V(\sigan)$, which can be equivalently re-written as in \eqref{re-express-V}. We analyze the attacker's equilibrium strategy set in each regime subsequently: 
\begin{enumerate}[label=(\alph*)]
\item Type I regimes $\regimei$:
\begin{itemize}
\item $i=0$:\\
Since $\ca > C_{(1)}-\Czero$, we must have $\Czero > \Ce-\ca$ for any $\e \in \Ebar$. There is no vulnerable facility, and thus $\siganwe(\emptyset)=1$. 

\item $i=1, \dots, \Ebarp$:\\
Since $\cd$ satisfies \eqref{regimei_notlast} or \eqref{regimei_last}, we obtain:
\begin{align}\label{sum_smaller_1}
\sum_{\e \in \cup_{k=1}^{i} \Ebar_{(k)}}\frac{\cd}{\Ce-\Czero}=\sum_{k=1}^{i} \frac{\cd \cdot \nnk}{\Cpk-\Czero} < 1
\end{align}
Therefore, the set of feasible attack strategies satisfying \eqref{multiple_bound}-\eqref{6d} is a non-empty set. We also know from Lemma \ref{only_attacked} that $\siganwe$ satisfies \eqref{sub:upper_bound}. Again from \eqref{regimei_notlast} or \eqref{regimei_last}, for any $k=1, \dots, i$, we have $\Cpk-\ca > \Czero$ and for any $k=i+1, \dots, \Ebarp$, we have $\Cpk-\ca < \Czero$. Since $\{C_{(k)}\}_{k=1}^K$ satisfy \eqref{order}, to maximize $V(\sigan)$ in \eqref{re-express-V}, the optimal solution must satisfy \eqref{multiple_bound}-\eqref{6d}. 
\end{itemize}
\item Type II regimes $\regimej$:
\begin{itemize}
\item $j=1$: From \eqref{regime_j_1}, we know that:
\begin{align}\label{larger_one}
1=\sum_{\e \in \Ebar_{(1)}} \siganwe(\e)< \frac{\cd E_{(1)}}{C_{(1)}-\Czero}.
\end{align}
Thus, the set of feasible attack strategies satisfying \eqref{sub:upper_one}-\eqref{sum_j_one} is a non-empty set. Additionally, from Lemma \ref{only_attacked}, we know that $\siganwe$ satisfies \eqref{sub:upper_one}. Since $C_{(1)}>\Cpk$ for any $k=2, \dots, \Ebarp$, and $C_{(1)}-\ca>\Czero$. From \eqref{re-express-V} and \eqref{larger_one}, we know that in equilibrium the attacker targets facilities in $\Ebar_{(1)}$ with probability 1. The set of strategies satisfying \eqref{sub:upper_one}-\eqref{sum_j_one} maximizes \eqref{re-express-V}, and thus is the set of attacker's equilibrium strategies.
\item $j=2, \dots, \Ebarp$: From \eqref{regime_j_rest}, we know that:
\begin{align*}
0 < 1-\sum_{k=1}^{j-1}\frac{\cd \cdot \nnk}{C_{(k)}-\Czero} < \frac{\cd \cdot E_{(j)}}{C_{(j)}-\Czero}.
\end{align*}
Thus, the set of feasible attack strategies satisfying \eqref{regime_k_1}-\eqref{regime_k_3} is a non-empty set. From Lemma \ref{only_attacked}, we know that $\siganwe$ satifies \eqref{regime_k_2}. Since $\{\Cpk\}_{k=1, \dots, j}$ satisfies the ordering in \eqref{order}, in order to maximize $V(\sigan)$ in \eqref{re-express-V}, $\siganwe$ must also satisfy \eqref{regime_k_1} and \eqref{regime_k_3}, and the remaining facilities are not targeted. 
\end{itemize}
\end{enumerate}

\vspace{0.2cm}
We next prove the defender's equilibrium security effort. By definition of Nash equilibrium, the probability vector $\pwe$ is induced by an equilibrium strategy if and only if it satisfies the following two conditions:
\begin{enumerate}
\item $\pwe$ is a best response to any $\siganwe \in \Siganwe$. 
\item Any attacker's equilibrium strategy is a best response to $\pwe$, i.e. the attacker has identical utilities for choosing any pure strategy in his equilibrium support set, and the utility is no less than that of any other pure strategies. 
\end{enumerate} 
Note that in both conditions, we require $\pwe$ to be a best response to \emph{any} attacker's equilibrium strategy. This is because given any $\siganwe \in \Siganwe$, $\(\pwe, \siganwe\)$ is an equilibrium strategy profile (Lemma \ref{zero_sum}). 
We now check these conditions in each regime:
\begin{enumerate}[label=(\alph*)]
\item Type I regimes $\regimei$:
\begin{itemize}
\item If $i=0$:\\
Since $\siganwe(\e)=0$ for any $\e \in \E$. From Lemma \ref{only_attacked}, the best response of the defender is $\pewe=0$ for any $\e \in \E$. 
\item If $i=1, \dots, \Ebarp$:\\
From Lemma \ref{only_attacked}, we know that $\pewe=0$ for any $\e \in \E \setminus \(\cup_{k=1}^{i} \Ebar_{(k)}\)$. Since $\siganwe(\emptyset)>0$, $\pewe$ must ensure that the attacker's utility of choosing any facility $\e \in \cup_{k=1}^{i}\Ebar_{(k)}$ is identical to that of choosing no attack $\emptyset$. Consider any $\e \in \cup_{k=1}^{i}\Ebar_{(k)}$:
\begin{alignat*}{2}
&&\Ua(\pwe, \e)&=\Ua(\pwe, \emptyset),\\ 
\stackrel{\eqref{Ua_rewrite}}{\Rightarrow} \quad &&\pewe \(\Czero-\ca\)+(1-\pewe) \(\Ce-\ca\)&=\Czero,\\
\Rightarrow \quad &&\pewe&=\frac{\Ce-\ca-\Czero}{\Ce-\Czero}, \quad \forall \e \in \cup_{k=1}^{i} \Ebar_{(k)}. 
\end{alignat*}
For any $\ebarnew \in \E \setminus \(\cup_{k=1}^{i}\Ebar_{(k)}\)$, since $\pwe_{\ebarnew}=0$, the attacker receives utility $C_{\ebarnew}-\ca$ by targeting $\ebarnew$, which is lower than $\Czero$. Therefore, $\pwe$ in \eqref{regime_last_sub}-\eqref{regime_last_zero} satisfies both conditions (1) and (2). $\pwe$ is the unique equilibrium strategy.
\end{itemize}
\item Type II regimes $\regimej$:
\begin{itemize}
\item If $j=0$:\\
Consider an attacker's strategy $\sigan$ such that:
\begin{align*}
\sigan(\e)&=\frac{1}{E_{(1)}}, \quad \forall \e \in \Ebar_{(1)}, \\
\sigan(\e)&=0, \quad \forall \e \in \E \setminus \Ebar_{(1)}. 
\end{align*}
Since $\cd$ satisfies \eqref{regime_j_1}, we know that $\frac{1}{E_{(1)}}< \frac{\cd}{C_{(1)}-\Czero}$. One can check that $\sigan$ satisfies \eqref{sub:upper_one}-\eqref{sum_j_one}, and thus $\sigan \in \Siganwe$. Therefore, we know from Lemma \ref{only_attacked} that $\pewe=0$ for any $\e \in \E$. 

\item If $j=1, \dots, \Ebarp$:\\
Analogous to our discussion for $j=0$, the following is an equilibrium strategy of the attacker:
\begin{align*}
\siganwe(\e)&=\frac{\cd}{\Ce-\Czero}, \quad \forall \e \in \cup_{k=1}^{j-1}\Ebar_{(k)}, \\
\siganwe(\e)&=\frac{1}{E_{(j)}} \(1- \sum_{i=1}^{j-1} \frac{\cd \nnk}{\Cpk-\Czero}\), \quad \forall \e \in \Ebar_{(j)}, \\
\siganwe(\e)&=0, \quad \forall \e \in \E \setminus \(\cup_{k=1}^{j} \Ebar_{(k)}\).
\end{align*}
From Lemma \ref{only_attacked}, we immediately obtain that $\pewe=0$ for any $\e \in \E \setminus \(\cup_{k=1}^{j-1} \Ebar_{(k)}\)$. 

Furthermore, for any $\e \in \cup_{k=1}^{j-1}\Ebar_{(k)}$, the utility of the attacker in choosing $\e$ must be the same as the utility for choosing any facility in $\Ebar_{(j)}$, which is $C_{(j)}-\ca$. Therefore, for any $\e \in \cup_{k=1}^{j-1}\Ebar_{(k)}$, $\pwe$ satisfies:
\begin{alignat*}{3}
&& &&\Ua(\pwe, \e)&=C_{(j)}-\ca, \\ 
&&\stackrel{\eqref{Ua_rewrite}}{\Rightarrow} \quad &&\pewe \(\Czero-\ca\)+(1-\pewe) \(\Cpk-\ca\)&=C_{(j)}-\ca, \\
&&\stackrel{\text{     }}{\Rightarrow} \quad &&\pewe&=\frac{C_{(k)}-C_{(j)}}{C_{(k)}-\Czero}.
\end{alignat*}
Additionally, for any $\e \in \E \setminus \(\cup_{k=1}^{j} \Ebar_{(k)}\)$, the utility for the attacker targeting $\e$ is $\Ce-\ca$, which is smaller than $C_{(j)}-\ca$. Thus, both condition (1) and (2) are satisfied. $\pwe$ is the unique equilibrium security effort.
\end{itemize}
\end{enumerate}


%

%
\section{Proofs of Section \ref{sequential_section}}\label{proof_sequential}
\noindent\emph{Proof of Lemma \ref{best_response_sequential}.}
For any non-vulnerable facility $\e$, the best response strategy $\sigas$ must be such that $\sigas(\e, \psss)=0$ for any $\psss$. 

Now consider any $\e \in \{\E| \Ce-\ca>\Czero\}$. If $\pess > \pebar$, then we can write:
\begin{align}\label{emptyset_dominated}
\Ua(\psss, \e)=\pess \Czero+(1-\pess) \Ce- \ca < \Czero=\Ua(\psss, \emptyset).
\end{align}
That is, the attacker's expected utility of targeting the facility $\e$ is less than the expected utility of no attack. Thus, in any attacker's best response, $\sigas(\e, \psss)=0$ for any such facility $\e$. Additionally, if $\pess = \pebar$, then $\Ua(\e, \psss)=\Ua(\emptyset, \psss)$, i.e. the utility of targeting such facility is identical with the utility of choosing no attack, and is higher than that of any other pure strategies. Hence, the set of best response strategies of the attacker is $\Delta(\Erho \cup\{\emptyset\})$, where $\Erho$ is the set defined in \eqref{Erho}. 

Otherwise, if there exists a facility $\e \in \{\E| \Ce-\ca>\Czero\}$ such that $\pess <\pebar$, then we obtain:
\[\Ua(\psss, \e)=\pess \Czero+(1-\pess) \Ce- \ca > \Czero=\Ua(\psss, \emptyset).\]
Thus, no attack cannot be chosen in any best response strategy, which implies that the attacker chooses to attack with probability 1. Finally, $\Emax$ is the set of facilities which incur the highest expected utility for the attacker given $\psss$, thus $BR(\psss) = \Delta(\Emax)$. \qed 

\vspace{0.3cm}
\noindent\emph{Proof of Lemma \ref{zero_or_one}.} 
We first prove that the total attack probability is either 0 or 1 in any SPE. We discuss the following three cases separately: 
\begin{itemize}
\item There exists at least one single facility $\e \in \{\Ebar | \Ce-\ca>\Czero\}$ such that $\psewe<\pebar$.\\
Since $\sigaswe(\pswe) \in BR(\pswe)$, from Lemma \ref{best_response_sequential}, we know that $\sum_{\e \in \Ebar} \sigaswe(\e, \pswe)=1$. 

\item For all $\e \in \{\Ebar | \Ce-\ca>\Czero\}$, $\psewe> \pebar$, i.e. the set $\Erho$ in \eqref{Erho} is empty. \\
Since $\sigaswe(\pswe) \in BR(\pswe)$, from Lemma \ref{best_response_sequential}, we know that no edge is targeted in SPE, i.e. $\sum_{\e \in \Ebar} \sigaswe(\e, \pswe)=0$. 

\item For all $\e \in \{\Ebar | \Ce-\ca>\Czero\}$, $\psewe\geq  \pebar$, and the set $\Erho$ in \eqref{Erho} is non-empty. \\
For the sake of contradiction, we assume that in SPE, there exists a facility $\e \in \Erho$ such that $\sigaswe(\e, \pswe)>0$, i.e. $\sigaswe(\emptyset, \pswe)<1$. Then, we can write $\Ud(\pswe, \sigaswe(\pswe))$ as follows: 
\begin{align}\label{Ud_indifferent}
\Ud(\pswe, \sigaswe(\pswe))=-\Czero-(1-\sigaswe(\emptyset, \pswe)) \ca -\(\sum_{\e \in \Ebar} \psewe\) \cd.  
\end{align}
Now, consider $\psssp$ as follows:
\begin{alignat*}{2}
\pessp&=\psewe+\epsilon> \pebar, &&\quad \forall \e \in \Erho, \\
\pessp&=\psewe=0, &&\quad \forall \e \in \E \setminus \Erho,
\end{alignat*}
where $\epsilon$ is a sufficiently small positive number. Given such a $\psssp$, we know from Lemma \ref{best_response_sequential} that the unique best response is $\sigas(\emptyset, \psssp)=1$. Therefore, the defender's utility is given by:
\begin{align*}
\Ud(\psssp, \sigas(\psssp))&=-\Czero - \(\sum_{\e \in \E} \pessp\) \cd.
\end{align*}
Additionally, 
\begin{align*}
\Ud(\psssp, \sigas(\psssp))-\Ud(\pswe, \sigas(\pswe)) = (1-\sigas(\emptyset, \pswe)) \ca -\epsilon \cd |\Erho|.
\end{align*}
Since $\epsilon$ is sufficiently small and $\sigas(\emptyset, \pswe)<1$, we obtain that $\Ud(\psssp, \sigas(\psssp))> \Ud(\pswe, \sigas(\pswe))$. Therefore, $\pswe$ cannot be a SPE. We can conclude that in this case, the attacker chooses not to attack with probability 1. 
\end{itemize}
\vspace{0.3cm}
We next show that in any SPE, the defender's security effort on each vulnerable facility $\e$ is no higher than the threshold $\pebar$ defined in \eqref{pebar}.  Assume for the sake of contradiction that there exists a facility $\ebar \in \{\Ebar|\Ce-\ca>\Czero\}$ such that $\pebarss >\pebarbar$. 
We discuss the following two cases separately:\\
\begin{itemize}
\item The set $\ebarp \in \{\Ebar|\Ce-\ca>\Czero, \pess<\pebar\}$ is non-empty. We know from Lemma \ref{best_response_sequential} that $BR(\psss)=\Delta(\Emax)$, where the set $\Emax$ in \eqref{Emax} is the set of facilities which incur the highest utility for the attacker. Clearly, $\Emax \subseteq \{\Ebar|\Ce-\ca>\Czero, \pess<\pebar\}$, and hence $\ebar \notin \Emax$. 

We consider $\psssp$ such that $\pebarssp=\pebarss-\epsilon$, where $\epsilon$ is a sufficiently small positive number, and $\pessp=\pess$ for any other facilities. 
%
Then $\pebarssp> \pebarbar$ still holds, and  the set $\Emax$ does not change. The attacker's best response strategy remains to be $BR(\psssp)=\Delta(\Emax)$. Hence, the utility of the defender given $\psssp$ increases by $\epsilon \cd$ compared to that given $\psss$, because the expected usage cost $\mathbb{E}_{\sigma}[C]$ does not change, but the expected defense cost decreases by $\epsilon \cd$. 
Thus, such $\psss$ cannot be the defender's equilibrium effort. 

\item For all $\e \in \{\Ebar|\Ce-\ca>\Czero\}$, $\pess \geq  \pebar$. We have already argued that $\sigaswe(\emptyset, \psss)=1$ in this case. Since the defense cost $\cd>0$, if there exists any $\e$ such that $\pess>\pebar$, then by decreasing the security effort on $\e$, the utility of the defender increases. Therefore, such $\psss$ cannot be an equilibrium strategy of the defender.
\end{itemize}
From both cases, we can conclude that for any $\e \in \{\Ebar|\Ce-\ca>\Czero\}$, $\psewe \leq \pebar$

\vspace{0.3cm}
Finally, any non-vulnerable facilities $\e \in \E \setminus \{\E|\Ce-\ca>\Czero\}$ will not be targeted, hence we must have $\psewe=0$. 
\qed

\vspace{0.3cm}

\noindent\emph{Proof of Lemma \ref{comparison_lemma}.}
We first show that the threshold $\cdtil(\ca)$ as given in \eqref{cdtil} is a well-defined function of $\ca$. Given any $0\leq \ca < C_{(1)}-\Czero$, there is a unique $i \in \{1, \dots, K\}$ such that $C_{(i+1)}-\Czero \leq \ca< C_{(i)}-\Czero$. Now, we need to show that there is a unique $j \in \{1, \dots, \i\}$ such that $\frac{\sum_{k=j+1}^{i}E_{(k)}}{\sum_{k=1}^{i} \frac{E_{(k)}}{\Cpk-\Czero}}\leq \ca<\frac{\sum_{k=j}^{i}E_{(k)}}{\sum_{k=1}^{i} \frac{E_{(k)}}{\Cpk-\Czero}}$ (or $0\leq \ca<\frac{E_{(i)}}{\sum_{k=1}^{i} \frac{E_{(k)}}{\Cpk-\Czero}}$ if $j=i$). Note that functions $\{\cdij\}_{j=1}^{\i}$ are defined on the range $\left[0, ~\frac{\sum_{k=1}^{i}E_{(k)}}{\sum_{k=1}^{i} \frac{E_{(k)}}{\Cpk-\Czero}}\right]$. Since $\{\Cpk\}_{k=1}^{i}$ satisfies \eqref{order}, we have: 
\begin{align*}
\frac{\sum_{k=1}^{i}E_{(k)}}{\sum_{k=1}^{i} \frac{E_{(k)}}{\Cpk-\Czero}} \geq \frac{\sum_{k=1}^{i}E_{(k)}}{\frac{1}{C_{(i)}-\Czero} \sum_{k=1}^{i} E_{(k)}}=C_{(i)}-\Czero.
\end{align*}
Hence, for any $C_{(i+1)}-\Czero \leq \ca< C_{(i)}-\Czero$, the value $\cdtil(\ca)$ is defined as $\cdij(\ca)$ for a unique $j \in \{1, \dots, i\}$. Therefore, we can conclude that for any $0 \leq \ca< C_{(1)}-\Czero$, $\cdtil(\ca)$ is a well-defined function.

We next show that $\cdtil(\ca)$ is continuous and strictly increasing in $\ca$. Since for any $i=1, \dots, K$, and any $j=1, \dots, i$, the function $\cdij(\ca)$ is continuous and strictly increasing in $\ca$, $\cdtil(\ca)$ must be piecewise continuous and strictly increasing in $\ca$. It remains to be shown that $\cdtil(\ca)$ is continuous at $\ca \in \left\{C_{(i)}-\Czero\right\}_{\i=2}^{K} \cup \left\{\frac{\sum_{k=j}^{i}E_{(k)}}{\sum_{k=1}^{i} \frac{E_{(k)}}{\Cpk-\Czero}}\right\}_{j=1, \dots, \i, \i=1, \dots K}$.

We now show that for any $i=2, \dots, K$, $\cdtil(\ca)$ is continuous at $C_{(i)}-\Czero$. Consider $\ca=C_{(i)}-\Czero-\epsilon$ where $\epsilon$ is a sufficiently small positive number. There is a unique $\jhat \in \{1, \dots, i\}$ such that $\cdtil(\ca)=\cd^{i\jhat}(\ca)$. We want to argue that $\jhat \neq \i$: 
\begin{alignat*}{2}
 &&\ca \cdot \(\sum_{k=1}^{i} \frac{E_{(k)}}{\Cpk-\Czero}\)&=\(C_{(i)}-\Czero-\epsilon\) \cdot \(\sum_{k=1}^{i} \frac{E_{(k)}}{\Cpk-\Czero}\)\\
 && &= E_{(i)}+\sum_{k=1}^{i-1}\frac{\(C_{(i)}-\Czero\) E_{(k)}}{C_{(k)}-\Czero}-\epsilon \(\sum_{k=1}^{i}\frac{E_{(k)}}{C_{(k)}-\Czero}\)> E_{(i)}, \\
\Rightarrow &&\quad \ca&=C_{(i)}-\Czero-\epsilon > \frac{E_{(i)}}{\sum_{k=1}^{i} \frac{E_{(k)}}{\Cpk-\Czero}}
\end{alignat*}
Thus, $\jhat \in \{1, \dots, i-1\}$, and from \eqref{cdtil}, $\frac{\sum_{k=\jhat+1}^{i}E_{(k)}}{\sum_{k=1}^{i} \frac{E_{(k)}}{\Cpk-\Czero}}\leq C_{(i)}-\Czero-\epsilon <\frac{\sum_{k=\jhat}^{i}E_{(k)}}{\sum_{k=1}^{i} \frac{E_{(k)}}{\Cpk-\Czero}}$. Since $\epsilon$ is a sufficiently small positive number, we have: 
\begin{alignat*}{2}
&&\sum_{k=\jhat+1}^{i}E_{(k)} &\leq \(\sum_{k=1}^{i} \frac{E_{(k)}}{\Cpk-\Czero}\) \cdot \(C_{(i)}-\Czero-\epsilon\)\\
&& &=E_{(i)}+\sum_{k=1}^{i-1}\frac{\(C_{(i)}-\Czero\) E_{(k)}}{C_{(k)}-\Czero}-\epsilon  \(\sum_{k=1}^{i} \frac{E_{(k)}}{\Cpk-\Czero}\)\\
\Rightarrow && \quad \sum_{k=\jhat+1}^{i-1}E_{(k)} &\leq  \sum_{k=1}^{i-1}\frac{\(C_{(i)}-\Czero\) E_{(k)}}{C_{(k)}-\Czero}+ \epsilon \(\sum_{k=1}^{i-1} \frac{E_{(k)}}{\Cpk-\Czero}\)\\
\Rightarrow && \quad \frac{\sum_{k=\jhat+1}^{i-1}E_{(k)}}{\sum_{k=1}^{i-1}\frac{E_{(k)}}{\Cpk-\Czero}} &\leq C_{(i)}-\Czero+\epsilon.
\end{alignat*}
Analogously, we can check that $ C_{(i)}-\Czero+\epsilon<\frac{\sum_{k=\jhat}^{i-1}E_{(k)}}{\sum_{k=1}^{i-1} \frac{E_{(k)}}{\Cpk-\Czero}}$. Hence, from \eqref{cdtil}, when $\ca=C_{(i)}-\Czero+\epsilon$, we have $\cdtil(\ca)=\cd^{i-1\jhat}(\ca)$. Then, 
\begin{align*}
\lim_{\ca \to \(C_{(i)}-\Czero\)^{-}}\cdtil(\ca)&=\lim_{\epsilon \to 0} \cd^{i\jhat}(C_{(i)}-\Czero-\epsilon)\\
&\stackrel{\eqref{cdij}}{=}\frac{C_{(\jhat)}-\Czero}{\(C_{(\jhat)}-\Czero\) \cdot \(\sum_{k=1}^{\jhat-1} \frac{E_{(k)}}{\Cpk-\Czero}\) + \sum_{k=\jhat}^{i-1} E_{(k)}-\sum_{k=1}^{i-1} \frac{\ca E_{(k)}}{\Cpk-\Czero}}\\
&=\lim_{\epsilon \to 0} \cd^{i-1\jhat}(C_{(i)}-\Czero+\epsilon)=\lim_{\ca \to \(C_{(i)}-\Czero\)^{+}}\cdtil(\ca).
\end{align*}
Thus, $\cdtil(\ca)$ is continuous at $C_{(i)}-\Czero$ for any $i=2, \dots, K$. 

For any $i=1, \dots, K$, we next show that $\cdtil(\ca)$ is continuous 
at $\ca=\frac{\sum_{k=j}^{i}E_{(k)}}{\sum_{k=1}^{i} \frac{E_{(k)}}{\Cpk-\Czero}}$ for $j=1, \dots, i$: 
\begin{align*}
\lim_{\ca \to \(\frac{\sum_{k=j}^{i}E_{(k)}}{\sum_{k=1}^{i} \frac{E_{(k)}}{\Cpk-\Czero}}\)^{-}}\cdtil(\ca)&=\cdij\(\frac{\sum_{k=j}^{i}E_{(k)}}{\sum_{k=1}^{i} \frac{E_{(k)}}{\Cpk-\Czero}}\)=\(\sum_{k=1}^{j-1} \frac{E_{(k)}}{\Cpk-\Czero}\)^{-1}\\
&=\cd^{i(j-1)}\(\frac{\sum_{k=j}^{i}E_{(k)}}{\sum_{k=1}^{i} \frac{E_{(k)}}{\Cpk-\Czero}}\)=\lim_{\ca \to \(\frac{\sum_{k=j}^{i}E_{(k)}}{\sum_{k=1}^{i} \frac{E_{(k)}}{\Cpk-\Czero}}\)^{+}}\cdtil(\ca). 
\end{align*}


Hence, we can conclude that $\cdtil(\ca)$ is continuous and strictly increasing in $\ca$. 

Additionally, for any $i=1, \dots, \Ebarp$, consider any $\ca$ such that $C_{(i+1)}-\Czero<\ca\leq C_{(i)}-\Czero$ (or $0<\ca\leq C_{(\Ebarp)}-\Czero$ if $i=\Ebarp$), then for any $j=1, \dots, i$, we have:
\begin{align*}
\cdij(\ca)& \stackrel{\eqref{cdij}}{=}\frac{C_{(j)}-\Czero}{\(C_{(j)}-\ca-\Czero\) \cdot \(\sum_{k=1}^{j-1} \frac{E_{(k)}}{\Cpk-\Czero}\) + \sum_{k=j}^{i} \frac{\(\Cpk-\ca-\Czero\) E_{(k)}}{\Cpk-\Czero}}\\
&>\frac{C_{(j)}-\Czero}{\(C_{(j)}-C_{(i+1)}\) \cdot \(\sum_{k=1}^{j-1} \frac{E_{(k)}}{\Cpk-\Czero}\) + \sum_{k=j}^{i} \frac{\(\Cpk-C_{(i+1)}\) E_{(k)}}{\Cpk-\Czero}}\\
&= \frac{C_{(j)}-\Czero}{\(C_{(j)}-C_{(i+1)}\) \cdot \(\sum_{k=1}^{i} \frac{E_{(k)}}{\Cpk-\Czero}\)} \\
&\stackrel{\eqref{order}}{>} \(\sum_{k=1}^{i} \frac{E_{(k)}}{\Cpk-\Czero}\)^{-1}\\
&\stackrel{\eqref{cd_accurate}}{=}\cdbar.
\end{align*}
Therefore, for any $0<\ca<C_{(1)}-\Czero$, we have:
\begin{align}\label{comparison}
\cdtil(\ca)\stackrel{\eqref{cdtil}}{\geq} \min_{j=1, \dots, i}\cdij(\ca)>\cdbar,
\end{align}
Finally, if $\ca=0$, then we know that $\cdtil(0)=\cd^{KK}(0)$. From \eqref{cdij}, we can check that $\cd^{KK}(0)=\(\sum_{k=1}^{K} \frac{E_{(k)}}{\Cpk-\Czero}\)^{-1}=\bar{\cd}(0)$. If $\ca$ approaches $C_{(1)}-\Czero$, then $\cdtil(\ca)=\cd^{11}(\ca)$, and we have:
\begin{align*}
\lim_{\ca \to C_{(1)}-\Czero}\cdtil(\ca)\stackrel{\eqref{cdij}}{=}\lim_{\ca \to C_{(1)}-\Czero} \frac{C_{(1)}-\Czero}{ E_{(1)}-\frac{\ca E_{(1)}}{C_{(1)}-\Czero}} = +\infty
\end{align*}
\qed 

\vspace{0.3cm}

We define the partition as:
\begin{align}\label{partition}
\mathcal{P}\deleq \left\{\left\{ \regimei\right\}_{i=0}^{\Ebarp}, \left\{\regimeij\right\}_{j=1, \dots, \i, i=1, \dots, \Ebarp,} \right\}, 
\end{align}
where $\left\{\regimei\right\}_{i=0}^{K}$ are type I regimes in the normal form game defined in \eqref{regimei_first}-\eqref{regimei_last}, and $\regimeij$ is the set of $\(\cd, \ca\)$, which satisfy:
\begin{subequations}\label{partition}
\begin{align}
\cd &\in \left\{
\begin{array}{ll}
\(\(\frac{E_{(1)}}{C_{(1)}-\Czero}\)^{-1}, +\infty\), & \quad \text{if $j=1$,} \\
\(\(\sum_{k=1}^{j} \frac{E_{(k)}}{\Cpk-\Czero}\)^{-1}, \(\sum_{k=1}^{j-1} \frac{E_{(k)}}{\Cpk-\Czero}\)^{-1}\), & \quad \text{if $j=2, \dots, \Ebarp$,}
\end{array}
\right.\label{cd_partition}
\\
\ca &\in \left\{
\begin{array}{ll}
\(C_{(i+1)}-\Czero, C_{(i)}-\Czero\), & \quad \text{if $i=1, \dots, \Ebarp-1$,} \\
\(0, C_{(\Ebarp)}-\Czero\), & \quad \text{if $i=\Ebarp$,}
\end{array}
\right.\label{ca_partition}
\end{align}
\end{subequations}
We can check that sets in $\mathcal{P}$ are disjoint, and cover the whole space of $\(\cd, \ca\)$. Lemma \ref{sequential_type1} characterizes SPE in sets $\left\{ \regimei\right\}_{i=0}^{\Ebarp}$, and Lemma \ref{type_2_sequential} characterizes SPE in sets $\left\{\regimeij\right\}_{i=1, j=1}^{i=\Ebarp, j=i}$.
\begin{lemma}\label{sequential_type1}
In $\games$, for any $\(\ca, \cd\)$ in the set $\regimei$, where $i=0, \dots, \Ebarp$:
\begin{itemize}
\item If $i=0$, then SPE is as given in \eqref{SPE_i_0}.
\item If $i=1, \dots, \Ebarp$: then SPE is as given in \eqref{SPE_i}.
\end{itemize}

\end{lemma}

\emph{Proof of Lemma \ref{sequential_type1}.}
\begin{itemize}
\item If $i=0$:\\
The set of vulnerable facilities $\{\Ebar|\Ce-\ca>\Czero\}$ is empty. Thus, $\sigaswe(\emptyset, \psss)=1$, and $\psewe=0$ for all $\e \in \E$.

\item For any $i=1, \dots, \Ebarp$:\\
The set of vulnerable facilities is $\cup_{k=1}^{i} \Ebar_{(k)}$. From Lemma \ref{zero_or_one}, we have already known that for any $\e \in \cup_{k=1}^{i} \Ebar_{(k)}$, $\psewe \leq \pebar$. Assume for the sake of contradiction that there exists a facility $\ebar \in \cup_{k=1}^{i} \Ebar_{(k)}$ such that $\pebarss <\pebarbar$. From Lemma \ref{best_response_sequential}, we know that $\sigaswe(\emptyset, \psss)=0$, and $BR(\psss)=\Delta(\Emax)$, where $\Emax$ is in \eqref{Emax}. Clearly, $\Emax \subseteq \cup_{k=1}^{i} \Ebari$. We define $\lambda$ as follows:
\begin{align*}
\lambda &=\max_{\e \in  \cup_{k=1}^{i} \Ebari} \left\{\pess \Czero+ (1-\pess) \Ce\right\}=\pess \Czero+ (1-\pess) \Ce, \quad \forall \e \in \Emax.
\end{align*}
The utility of the defender can be written as:
\begin{align*}
\Ud(\psss, \sigaswe(\psss))=-\lambda-\(\sum_{\e \in \E} \pess \)\cdot \cd.
\end{align*}

We now consider $\psssp$ as follows:
\begin{alignat*}{2}
\pessp&=\pess+\frac{\epsilon}{\Ce-\Czero}, &&\quad \forall \e \in \Emax, \\
\pessp&=\pess, &&\quad \forall \e \in \E \setminus\Emax,
\end{alignat*}
where $\epsilon$ is a sufficiently small positive number. Under this deviation, we can check that the set $\Emax$ does not change, but $\lambda$ changes to $\lambda-\epsilon$. Therefore, the defender's utility can be written as:
\begin{align*}
&\Ud(\psssp, \sigas(\psssp))=-\lambda+\epsilon -\(\sum_{\e \in \E} \psssp_\e \)\cdot \cd=-\lambda+\epsilon-\(\sum_{\e \in \E} \psss_\e \)\cdot \cd -\sum_{\e \in \Emax}\frac{\epsilon \cd}{\Ce-\Czero}\\
=&\Ud(\psss, \sigas(\psss))+\epsilon \(1-\sum_{\e \in \Emax}\frac{ \cd}{\Ce-\Czero}\)\geq \Ud(\psss, \sigas(\psss))+\epsilon \(1-\sum_{\e \in \cup_{k=1}^{i}\Ebari}\frac{ \cd}{\Ce-\Czero}\)\\
=& \Ud(\psss, \sigas(\psss))+\epsilon \(1-\sum_{k=1}^{i}\frac{ \cd \nnk}{\Ce-\Czero}\)\stackrel{\eqref{regimei_notlast}}{>}\Ud(\psss, \sigas(\psss)).
\end{align*}
Therefore, such $\psss$ cannot be an equilibrium strategy profile. We thus know that $\pswe$ is as given in \eqref{SPE_i}. The attacker's equilibrium strategy can be derived from Lemmas \ref{best_response_sequential} and \ref{zero_or_one} directly. \qed
\end{itemize}


\begin{lemma}\label{type_2_sequential}
For $\(\ca, \cd\)$ in $\regimeij$, where $i=1, \dots, \Ebarp$, and $j=1, \dots, i$, there are two cases of SPE:
\begin{itemize}
\item If $\cd>\cdij$, where $\cdij$ is as given in \eqref{cdij}:
\begin{itemize}
\item If $j=1$, then SPE is as given in \eqref{SPE_j_1}.
\item If $j=2, \dots, i$, then SPE is as given in \eqref{SPE_j}.
\end{itemize} 
\item If $\cd<\cdij$, then the SPE is as given in \eqref{SPE_i}. 
\end{itemize}
\end{lemma}

\emph{Proof of Lemma \ref{type_2_sequential}.}
Consider cost parameters in the set $\regimeij$ defined in \eqref{partition}, where $i=1, \dots, \Ebarp$ and $j=1, \dots, i$. The set of vulnerable facilities is $\cup_{k=1}^{i} \Ebar_{(k)}$. From Lemma \ref{zero_or_one}, we know that the defender can either secure all vulnerable facilities $\e \in \cup_{k=1}^{i} \Ebar_{(k)}$ with the threshold effort $\pebar$ defined in \eqref{pebar}, or leave at least one vulnerable facility secured less than the threshold effort. We discuss the two cases separately: 
\begin{enumerate}
\item[(1)] If any $\e \in \cup_{k=1}^{i} \Ebar_{(k)}$ is secured with the threshold effort $\pebar$, then from Lemma \ref{zero_or_one}, we know that the total probability of attack is 0. The defender's utility can be written as: 
\begin{align}
\Ud(\pbar, \sigaswe(\pbar))=-\Czero-\(\sum_{k=1}^{i} \frac{\(\Cpk-\ca-\Czero\) \cdot E_{(k)}}{\Cpk-\Czero}\) \cdot \cd. \label{ud_case2}
\end{align}
\item[(2)] If the set $\{\E|\Ce-\ca>\Czero, \quad \pess < \pebar\}$ is non-empty, then we define $\rhoone$ as the set of feasible $\psss$ in this case. We denote $\psssone$ as the secure effort vector that incurs the highest utility for the defender among all $\psss \in \rhoone$. Then, $\psssone$ can be written as: 
\begin{align}
&\psssone \in \underset{\psss \in \rhoone}{\mathrm{argmax}}~ \Ud(\psss, \sigaswe(\psss))=\underset{\psss  \in \rhoone}{\mathrm{argmax}} \(-\mathbb{E}_{\(\psss, \sigaswe(\psss)\)}[C]-\(\sum_{\e \in \E} \pess\)\cdot \cd\) \notag \\
=&\underset{\psss  \in \rhoone}{\mathrm{argmax}} \(-\mathbb{E}_{\(\psss, \sigaswe(\psss)\)}[C]-\(\sum_{\e \in \E} \pess\)\cdot \cd +  \(\sum_{\e \in \E} \sigaswe(\e, \psss)\)\cdot \ca-  \(\sum_{\e \in \E} \sigaswe(\e, \psss)\)\cdot \ca \). \label{psssone}
\end{align}
We know from Lemma \ref{best_response_sequential} that $\sigaswe(\emptyset, \psss)=0$. Therefore, $\sum_{\e \in \E} \sigaswe(\e, \psss)=1$, and \eqref{psssone} can be re-expressed as:
\begin{align*}
\psssone &\in  \underset{\psss  \in \rhoone}{\mathrm{argmax}} \(-\mathbb{E}_{\(\psss, \sigaswe(\psss)\)}[C]-\(\sum_{\e \in \E} \pess\)\cdot \cd +  \(\sum_{\e \in \E} \sigaswe(\e, \psss)\)\cdot \ca-  \ca \)\\
&=\underset{\psss  \in \rhoone}{\mathrm{argmax}} \(-\mathbb{E}_{\(\psss, \sigaswe(\psss)\)}[C]-\(\sum_{\e \in \E} \pess\)\cdot \cd +  \(\sum_{\e \in \E} \sigaswe(\e, \psss)\)\cdot \ca \). 
\end{align*}
Since in equilibrium, the attacker chooses the best response strategy, we have:
\begin{align}\label{zero_sum_again}
\mathbb{E}_{\(\psss, \sigaswe(\psss)\)}[C]- \(\sum_{\e \in \E} \sigaswe(\e, \psss)\)\cdot \ca =\max_{\sigas \in \Delta(\Sa)} \(\mathbb{E}_{\(\psss, \sigas\)}[C]- \(\sum_{\e \in \E} \sigas(\e)\) \cdot \ca\).
\end{align}
Hence, $\psssone$ can be re-expressed as:
\begin{align*}
\psssone&\stackrel{\eqref{zero_sum_again}}{=} \underset{\psss  \in \rhoone}{\mathrm{argmax}} \(-\max_{\sigas \in \Delta(\Sa)} \(\mathbb{E}_{\(\psss, \sigas\)}[C]- \(\sum_{\e \in \E} \sigas(\e)\) \cdot \ca\)-\(\sum_{\e \in \E} \pess\)\cdot \cd \)\\
&=\underset{\psss  \in \rhoone}{\mathrm{argmax}} \(-\max_{\sigas \in \Delta(\Sa)} \(\mathbb{E}_{\(\psss, \sigas\)}[C]-\(\sum_{\e \in \E} \sigas(\e)\) \cdot \ca +\(\sum_{\e \in \E} \pess\)\cdot \cd \)\)\\
&=\underset{\psss  \in \rhoone}{\mathrm{argmax}}\min_{\sigas \in \Delta(\Sa)} \(-\mathbb{E}_{\(\psss, \sigas\)}[C]+\(\sum_{\e \in \E} \sigas(\e)\) \cdot \ca -\(\sum_{\e \in \E} \pess\)\cdot \cd \)\\
&\stackrel{\eqref{zero_utility_defend}}{=} \underset{\psss  \in \rhoone}{\mathrm{argmax}}\min_{\sigas \in \Delta(\Sa)}  ~ \Udzero(\psss, \sigas).
\end{align*}
Therefore, $\psssone$ is the defender's equilibrium strategy in the zero sum game, which is identical to the equilibrium strategy in the normal form game (recall Lemma \ref{zero_sum}). From Theorem \ref{attacker_strategy}, when $\ca$ and $\cd$ are in $\regimeij$, $\psssone$ is in \eqref{SPE_j} (or \eqref{SPE_j_1} if $j=1$).  
The defender's utility in this case is:
\begin{align}
\Ud(\psssone, \sigaswe(\psssone))=-C_{(j)}-\(\sum_{k=1}^{j-1} \frac{\(\Cpk-C_{(j)}\) \cdot E_{(k)}}{\Cpk-\Czero}\) \cdot \cd. \label{ud_case1}
\end{align}
\end{enumerate}
Finally, by comparing $\Ud$ in \eqref{ud_case1} and \eqref{ud_case2}, we can check that if $\cd>\cdij$, then $\Ud(\psssone, \sigaswe(\psssone))> \Ud(\pbar, \sigaswe(\pbar))$. Thus, SPE is in \eqref{SPE_j} (or \eqref{SPE_j_1} if $j=1$). If $\cd<\cdij$, then $\Ud(\psssone, \sigaswe(\psssone))< \Ud(\pbar, \sigaswe(\pbar))$, and SPE is in \eqref{SPE_i}. \qed

\vspace{0.3cm}

\noindent\emph{Proof of Theorem \ref{theorem:SPE}.}
\begin{itemize}
\item[(a)] Type $\typeone$ regimes $\regimesi$:
\begin{itemize}
\item If $i=0$:\\
There is no vulnerable facility. Therefore, the attacker chooses not to attack with probability 1, and the defender does not secure any facility. SPE is as given in \eqref{SPE_i_0}.
\item If $i=1, \dots, \Ebarp$:\\
Consider any $C_{(i+1)}-\Czero < \ca < C_{(i)}-\Czero$. From Lemma \eqref{comparison_lemma}, we know that $\cdtil(\ca)> \cdbar$, where $\cdtil(\ca)$ is defined in \eqref{cdtil} and $\cdbar$ is as defined in \eqref{cdbar}. From Lemma \ref{sequential_type1}, we know that SPE is as given in \eqref{SPE_i} for any $\cd< \cdbar$. 

It remains to be shown that for any $\cdbar \leq \cd < \cdtil(\ca)$, SPE is also as given in \eqref{SPE_i}. For any $C_{(i+1)}-\Czero \leq \ca <C_{(i)}-\Czero$, there is a unique $\jhat \in \{1, \dots, i\}$ such that $\frac{\sum_{k=\jhat+1}^{i}E_{(k)}}{\sum_{k=1}^{i} \frac{E_{(k)}}{\Cpk-\Czero}}\leq \ca<\frac{\sum_{k=\jhat}^{i}E_{(k)}}{\sum_{k=1}^{i} \frac{E_{(k)}}{\Cpk-\Czero}}$, and from \eqref{cdtil}, we have:
\begin{align*}
\cdtil(\ca)&=\cd^{i\jhat}(\ca)\geq \cd^{i\jhat}\(\frac{\sum_{k=\jhat+1}^{i}E_{(k)}}{\sum_{k=1}^{i} \frac{E_{(k)}}{\Cpk-\Czero}}\)\stackrel{\eqref{cdij}}{=}\(\sum_{k=1}^{\jhat} \frac{E_{(k)}}{\Cpk-\Czero}\)^{-1}, \\
\cdtil(\ca)&=\cd^{i\jhat}(\ca)<\cd^{i\jhat}\(\frac{\sum_{k=\jhat}^{i}E_{(k)}}{\sum_{k=1}^{i} \frac{E_{(k)}}{\Cpk-\Czero}}\)=\(\sum_{k=1}^{\jhat-1} \frac{E_{(k)}}{\Cpk-\Czero}\)^{-1}.
\end{align*}
Consider any $j=\jhat+1, \dots, i$, and any $\(\sum_{k=1}^{j} \frac{E_{(k)}}{\Cpk-\Czero}\)^{-1} \leq \cd <\(\sum_{k=1}^{j-1} \frac{E_{(k)}}{\Cpk-\Czero}\)^{-1}$, the cost parameters $(\ca, \cd)$ are in the set $\regimeij$ as defined in \eqref{partition}. Additionally, from our definition of $\jhat$, we know that $\ca>\frac{\sum_{k=j}^{i}E_{(k)}}{\sum_{k=1}^{i} \frac{E_{(k)}}{\Cpk-\Czero}}$. We now show that in $\regimeij$, $\cd<\cdij(\ca)$: 
\begin{align*}
\cdij(\ca)\stackrel{\eqref{cdij}}{>}\cdij\(\frac{\sum_{k=j}^{i}E_{(k)}}{\sum_{k=1}^{i} \frac{E_{(k)}}{\Cpk-\Czero}}\)=\(\sum_{k=1}^{j-1} \frac{E_{(k)}}{\Cpk-\Czero}\)^{-1}\stackrel{\eqref{partition}}{>}\cd.
\end{align*}
Hence, from Lemma \ref{type_2_sequential}, we know that for any $\(\sum_{k=1}^{i} \frac{E_{(k)}}{\Cpk-\Czero}\)^{-1}\leq \cd \leq \(\sum_{k=1}^{\jhat} \frac{E_{(k)}}{\Cpk-\Czero}\)^{-1}$, SPE is as given in \eqref{SPE_i}. For any $\(\sum_{k=1}^{\jhat} \frac{E_{(k)}}{\Cpk-\Czero}\)^{-1}<\cd< \cdtil(\ca)$, the cost parameters $(\ca, \cd)$ are in the set $\Lambda_i^{\jhat}$, and $\cd<\cdtil(\ca)=\cd^{i\jhat}(\ca)$. Again from Lemma \ref{type_2_sequential}, SPE is in \eqref{SPE_i}. 

Therefore, we can conclude that in regime $\regimesi$, SPE is in \eqref{SPE_i}. 
\end{itemize}
\item[(b)] Type $\typetwo$ regimes $\regimesj$, where $j=1, \dots, K$:\\
Since $\cdtil(\ca)$ is strictly increasing in $\ca$ and $\lim_{\ca \to C_{(1)}-\Czero}\cdtil(\ca)=+\infty$, we know that for any $\cd>0$, $\ca<\cdtil^{-1}(\cd)<C_{(1)}-\Czero$. Therefore, we can re-express $\widetilde{\Lambda}^1$ as follows: 
\begin{align}
\widetilde{\Lambda}^1&\stackrel{\eqref{regimej_constraint_1}}{=} \left\{\(\ca, \cd\) \left \vert   \ca< \cdtil^{-1}(\cd), ~  \cd >  \(\frac{E_{(1)}}{C_{(1)}-\Czero}\)^{-1} \right.\right\}\notag\\
&=\left\{\(\ca, \cd\) \left \vert  \cd> \cdtil(\ca), ~  \cd >  \( \frac{E_{(1)}}{C_{(1)}-\Czero}\)^{-1}, 0 \leq \ca \leq C_{(1)}-\Czero \right.\right\}\notag\\
& \stackrel{\eqref{partition}}{=}=\bigcup_{i=1}^K \(\regimeij \bigcap \left\{\(\ca, \cd\)|\cd> \cdtil(\ca)\right\}\).\label{re_express_regimesone}
\end{align}
For any $j=2, \dots, K$, if $\ca>C_{(j)}-\Czero$, then from Lemma \ref{comparison_lemma}, we have: 
\begin{align}\label{empty_set}
\cdtil(\ca)> \cdbar \stackrel{\eqref{cd_accurate}}{\geq} \(\sum_{k=1}^{j-1} \frac{E_{(k)}}{\Cpk-\Czero}\)^{-1}.
\end{align}
Therefore, for any $\cd<\(\sum_{k=1}^{j-1} \frac{E_{(k)}}{\Cpk-\Czero}\)^{-1}$, we know that $\ca<\cdtil^{-1}(\cd)<C_{(j)}-\Czero$. Analogous to \eqref{re_express_regimesone}, we re-express the set $\regimesj$ as follows:
\begin{align*}
\regimesj &\stackrel{\eqref{regimej_constraint}}{=} \left\{\(\ca, \cd\) \left \vert   \ca< \cdtil^{-1}(\cd), ~ \(\sum_{k=1}^{j} \frac{E_{(k)}}{\Cpk-\Czero}\)^{-1}\leq \cd < \(\sum_{k=1}^{j-1} \frac{E_{(k)}}{\Cpk-\Czero}\)^{-1} \right.\right\}\\
&\stackrel{\eqref{empty_set}}{=}\left\{\(\ca, \cd\) \left \vert  \begin{array}{l}
 \cd> \cdtil(\ca), ~ \(\sum_{k=1}^{j} \frac{E_{(k)}}{\Cpk-\Czero}\)^{-1}\leq \cd < \(\sum_{k=1}^{j-1} \frac{E_{(k)}}{\Cpk-\Czero}\)^{-1}, \\
 0 \leq \ca \leq C_{(j)}-\Czero\end{array} \right.\right\}\\
&\stackrel{\eqref{partition}}{=}\bigcup_{i=j}^K \(\regimeij \bigcap \{\(\ca, \cd\)|\cd> \cdtil(\ca)\}\).
\end{align*}
We next show that for any $j=1, \dots, K$, and any $i=j, \dots, K$, the set $\regimeij \cap \{\(\ca, \cd\)|\cd> \cdtil(\ca)\} \subseteq \regimeij \cap \{\(\ca, \cd\)|\cd>\cdij(\ca)\}$. Consider any cost parameters $\(\ca, \cd\)$ in the set $\regimeij\cap \{\(\ca, \cd\)|\cd> \cdtil(\ca)\}$, from \eqref{cdtil}, we can find $\jhat$ such that $\frac{\sum_{k=\jhat+1}^{i}E_{(k)}}{\sum_{k=1}^{i} \frac{E_{(k)}}{\Cpk-\Czero}}\leq \ca<\frac{\sum_{k=\jhat}^{i}E_{(k)}}{\sum_{k=1}^{i} \frac{E_{(k)}}{\Cpk-\Czero}}$, and $\cdtil(\ca)=\cd^{i\jhat}(\ca)$. We discuss the following three cases separately: 
\begin{itemize}
\item If $\jhat > j$, then we must have $\ca<\frac{\sum_{k=\jhat}^{i}E_{(k)}}{\sum_{k=1}^{i} \frac{E_{(k)}}{\Cpk-\Czero}}\leq\frac{\sum_{k=j+1}^{i}E_{(k)}}{\sum_{k=1}^{i} \frac{E_{(k)}}{\Cpk-\Czero}}$. Hence, from \eqref{cdij}, $\cdij(\ca)<\(\sum_{k=1}^{j} \frac{E_{(k)}}{\Cpk-\Czero}\)^{-1}$. From the definition of the set $\regimeij$ in \eqref{partition}, we know that $\cd>\cdij(\ca)$ in this set, and thus $\(\ca, \cd\) \in \regimeij\cap \{\(\ca, \cd\)|\cd> \cdij(\ca)\}$.
\item If $\jhat=j$, then we directly obtain that $\(\ca, \cd\) \in \regimeij\cap \{\(\ca, \cd\)|\cd> \cdij(\ca)\}$. 
\item If $\jhat<j$, then since $\ca \geq \frac{\sum_{k=\jhat+1}^{i}E_{(k)}}{\sum_{k=1}^{i} \frac{E_{(k)}}{\Cpk-\Czero}}$, from \eqref{cdij}, we have $\cdtil(\ca)=\cd^{i\jhat}(\ca)\geq \(\sum_{k=1}^{\jhat} \frac{E_{(k)}}{\Cpk-\Czero}\)^{-1} \geq \(\sum_{k=1}^{j-1} \frac{E_{(k)}}{\Cpk-\Czero}\)^{-1}$. From the definition of the set $\regimeij$ in \eqref{partition}, the set $\regimeij\cap \{\(\ca, \cd\)|\cd> \cdtil(\ca)\}$ is empty, and thus can be omitted. 
\end{itemize}
We can conclude from all three cases that $\regimeij \cap \{\(\ca, \cd\)|\cd> \cdtil(\ca)\} \subseteq \regimeij \cap \{\(\ca, \cd\)|\cd>\cdij(\ca)\}$. Therefore, from Lemma \ref{type_2_sequential}, SPE is in \eqref{SPE_j} (or \eqref{SPE_j_1} if $j=1$) in the regime $\regimesj$. 

\end{itemize}
\qed

\end{appendix}



\end{document}